\documentclass[mnsc,nonblindrev]{informs4rad}

\newif\ifblind
\blindfalse 

\newcommand{\revcolor}[1]{{\color{black}#1}}

\OneAndAHalfSpacedXI

\makeatletter
\newdimen\INFORMS@TITLEwidth
\INFORMS@TITLEwidth=\textwidth
\advance\INFORMS@TITLEwidth by 0.07in 

\def\theARTICLETITLE{%
  \HOOKtop
  \begin{Center}
    \vspace*{0pt}%
    \begin{minipage}{\INFORMS@TITLEwidth}
      \centering
      \TITLEfont\HD{24}{0}\theTITLE\HD{0}{15}%
    \end{minipage}%
  \end{Center}}
\makeatother

\usepackage[suppress]{color-edits}

\addauthor{FE}{blue}
\makeatletter
\renewcommand{\FEdelete}[1]{}
\makeatother

\usepackage{mathtools}

\usepackage{xcolor}
\usepackage{natbib}
\usepackage{mathrsfs}
\usepackage{dsfont}
\definecolor{cornellred}{rgb}{0.7, 0.11, 0.11}
\definecolor{maroon}{rgb}{0.52, 0, 0}
\definecolor{dgreen}{rgb}{0.0, 0.5, 0.0}
\definecolor{ballblue}{rgb}{0.13, 0.67, 0.8}
\definecolor{royalblue(web)}{rgb}{0.25, 0.41, 0.88}
\definecolor{bleudefrance}{rgb}{0.19, 0.55, 0.91}
\definecolor{royalazure}{rgb}{0.0, 0.22, 0.66}
\usepackage{hyperref}
\hypersetup{
	colorlinks = true,
	linkcolor=maroon,
	urlcolor=royalazure,
	citecolor=royalazure,
	linkbordercolor = {red}
}
\usepackage{cleveref}

\usepackage{enumitem}
\usepackage{xcolor}
\usepackage{booktabs} 
\usepackage{amsmath}
\usepackage{subcaption}
\usepackage{float}
\usepackage{tikz}
\usetikzlibrary{shapes.geometric, arrows.meta, positioning, calc}
\usetikzlibrary{calc}

\usepackage{csquotes}
\makeatletter
\renewenvironment*{displayquote}
  {\begingroup\setlength{\leftmargini}{0.3cm}\csq@getcargs{\csq@bdquote{}{}}}
  {\csq@edquote\endgroup}
\makeatother

\renewcommand{\qed}{$\hfill\square$}

\newenvironment{proofof}[1]{%
  \Trivlist
  \item[\hskip\labelsep {\it #1.}]\ignorespaces
}{\hfill \qed
\endTrivlist
\addvspace{0pt}
}

\newcommand{\prob}{\text{I\kern-0.15em P}}

\newcommand{\driverArrival}{\tilde{\lambda}}
\newcommand{\idleChurn}{\tilde{\eta}_{\text{I}}}
\newcommand{\notifChurn}{\tilde{\eta}_{\text{N}}}
\newcommand{\rideChurn}{\eta}
\newcommand{\driverTotalArrival}{\tilde{\lambda}}

\newcommand{\riderTotalArrival}{\lambda}
\newcommand{\rides}{\mathcal{R}}
\newcommand{\riders}{\mathcal{R}}
\newcommand{\drivers}{\mathcal{D}}

\usepackage[ruled,vlined,linesnumbered]{algorithm2e}

\SetCommentSty{mycommfont}

\DontPrintSemicolon
\SetAlgoVlined
\SetAlFnt{\small}
\SetAlCapFnt{\small}
\SetAlCapNameFnt{\small}
\SetAlCapHSkip{0pt}
\IncMargin{-\parindent}
\setcitestyle{authoryear}

\allowdisplaybreaks
\newtheorem{theorem}{Theorem}[section]
\usepackage{thm-restate}

\newtheorem{proposition}[theorem]{Proposition}

\newtheorem{definition}{Definition}[section]

\newtheorem{remark}[definition]{Remark}
\newtheorem{example}[definition]{Example}
	\newtheorem{assumption}[definition]{Assumption}

\newcommand{\xhdr}[1]{\noindent\textbf{#1}}

\newcommand{\RNcolor}[1]{{\textcolor{black}{#1}}}
\MANUSCRIPTNO{}
\begin{document}
\RUNTITLE{Non-Exclusive Notifications for Ride-Hailing at Lyft II: Simulations and Marketplace Analysis}
\TITLE{Non-Exclusive Notifications for Ride-Hailing at Lyft II: Simulations and Marketplace Analysis}
\RUNAUTHOR{Ekbatani et al.}

\ARTICLEAUTHORS{

\AUTHOR{Farbod Ekbatani, Rad Niazadeh}
\AFF{University of Chicago Booth School of Business, Chicago, IL\\
\EMAIL{fekbatan@chicagobooth.edu, rad.niazadeh@chicagobooth.edu}}

\AUTHOR{Mehdi Golari, Romain Camilleri, Titouan Jehl, Chris Sholley}
\AFF{Lyft, Inc.\\ \EGT\EMAIL{mehdig@lyft.com, rcamilleri@lyft.com, titouanj@lyft.com, chris@lyft.com}}

\AUTHOR{Matthew Leventi, Theresa Calderon, Angela Lam, Paul Havard Duclos}
\AFF{Lyft, Inc.\\ \EGT\EMAIL{mleventi@lyft.com, tcalderon@lyft.com, alam@lyft.com, paul.havardduclos@gmail.com}}

\AUTHOR{Tim Holland, James Koch, Shreya Reddy}
\AFF{Lyft, Inc.\\ \EGT\EMAIL{tholland@lyft.com, jkoch@lyft.com, sreddy@lyft.com}}

}




\ABSTRACT{
Ride-hailing platforms increasingly face uncertain driver acceptance, which makes traditional one-to-one \emph{exclusive dispatch} (ED) less efficient: rejections and timeouts force sequential retries and lengthen rider wait times, which in turn creates friction in the marketplace. \emph{Non-exclusive dispatch} (NED) mitigates this friction by broadcasting a request to multiple drivers in parallel. While NED can reduce latency, it introduces new design challenges---most notably, how to choose notification sets and how to resolve driver contention (when multiple drivers accept the same ride).

In this paper---the second in a two-part collaboration with Lyft---we develop a theoretically grounded framework to evaluate the long-run performance and marketplace effects of transitioning from ED to NED. We bridge theory and practice by combining (i) an optimization model that formulates NED as a constrained welfare maximization problem with (ii) large-scale discrete-event simulations on proprietary Lyft traces and (iii) a stylized macroscopic equilibrium model. Across simulation and equilibrium analysis, we find that NED improves key fulfillment metrics relative to ED: it reduces match time (and hence rider reneging) while increasing both the number and the average quality of completed matches. We also quantify the speed--quality trade-off between two common contention resolution rules, \emph{First-Accept} and \emph{Best-Accept}: First-Accept maximizes speed and throughput, whereas Best-Accept is required to maximize per-match quality. Finally, we show that slightly conservative notification heuristics can improve long-run efficiency by avoiding excessive locking of high-value drivers and preserving future availability.

}

\maketitle

\newpage
\section{Introduction}
\label{sec:intro}
Two-sided online platforms often clear supply and demand by repeatedly matching customers to service providers. In the early days of several of these platforms, one-to-one matching---that is, connecting a single customer with a single service provider---served as the backbone mechanism to match demand and supply. For example, in ride-hailing, dispatch typically runs in short cycles: the platform observes batches of unmatched ride requests and available drivers, computes matching decisions, and sends notifications to drivers to accept~\citep{xu2018orderdispatch,azagirre2024better}. Such cycle-based matchings are traditionally based on \emph{exclusive dispatch} (ED): each ride request is offered to a single driver. If that driver rejects or times out, the rider remains unmatched in that cycle and must be re-attempted in subsequent cycles~\citep{xu2018orderdispatch,zhang2017taxidispatch}. ED is operationally simple---it avoids conflicts by construction---but its efficiency heavily relies on the assumption that drivers have high acceptance rates at the time of notification~\citep{ozkan2020dynamic}.


In modern ride-hailing platforms, this assumption is increasingly untenable and is being challenged by the drivers. These drivers, as gig workers of the platforms, retain discretion over which trips to accept; Their acceptance behavior reflects heterogeneous preferences, information revealed at the time of the offer, and platform incentives~\citep{rheingans2019ridesharing,varma2021near,horner2021optimizing,chen2025grab}. Under ED, possible rejections are handled sequentially: a rejected request returns to the idle pool and is re-offered to another driver in a later cycle. This process repeats until the platform finds a driver who accepts the request. Although logically coherent, this procedure is structurally fragile. As rejections and timeouts become more frequent, sequential re-offering creates a bottleneck that increases rider waiting times and abandonment, ultimately deteriorating marketplace efficiency~\citep{castro2020matchingqueues,aveklouris2025matching}.



This paper---and our more algorithmic companion paper~\ifblind\cite{LyftI2026companion_Blind}\else\cite{LyftI2026companion_NonBlind}\fi---grew out of an academic–industry collaboration with Lyft, complementing a growing body of marketplace research grounded in ride-hailing platforms and industry partnerships~\citep{azagirre2024better,castillo2025matching,yan2025trading}. Our research collaboration is motivated by a specific operational shift in the dispatch pipeline to counteract the above mentioned issues: moving from exclusive one-to-one offers to \emph{non-exclusive dispatch} (NED), in which a ride request is sent to multiple nearby drivers in parallel~\citep{sun2020taxi,qin2025two_round,chen2023dynamic}.\footnote{This design is closely related to ``broadcasting'' or ``order-grabbing'' systems studied in transportation (where drivers choose whether to ``grab'' a broadcast request) and to ``supplier menu'' designs in peer-to-peer logistics and freight platforms (where the same task is assigned to a menu of suppliers). For example, see~\cite{liu2025recommend,chen2025grab,mofidi2019beneficial,ausseil2022supplier,horner2025increasing}.} By parallelizing notifications, the platform hedges against individual rejections and increases the probability of a timely successful match (without incurring the delays associated with sequential retries), improving rider waiting times and match quality. At the same time, higher fulfillment can expand drivers' earning opportunities while maintaining driver experience through safeguards used in practice.





However, moving from ``one-to-one'' to ``many-to-one'' notifications introduces new layers of (highly non-trivial) operational complexities, and in particular creates two tightly coupled design problems. First is \emph{notification packing}: selecting (disjoint) notification sets---which are combinatorial objects---subject to platform constraints. Second is \emph{contention resolution}: deciding who gets the ride when multiple drivers accept the same broadcast notification. We focus on two contention-resolution protocols that are common in practice.
\begin{itemize}
    \item \textit{First Acceptance (FA):} The platform assigns the ride to the first driver who responds. This models a fast dynamic, prioritizing speed but potentially sacrificing match quality (e.g., a closer driver might be slower to interact with the app than a distant one). 
    \smallskip
    
    \item \textit{Best Acceptance (BA):} The platform collects all responses and assigns the ride to the highest-scoring (e.g., the closest) driver among those who accepted. This prioritizes the quality of the match but introduces a mandatory waiting period to elicit responses from all drivers.
\end{itemize}
This choice reflects a fundamental trade-off between match speed and match quality, which mirrors similar tensions in other broadcast-offer environments---including food rescue and donation platforms, community first-responder dispatch, and crowdsourced delivery---where a recurring design fork is whether to commit immediately to the first responder or to wait and select the best among timely responders~\citep{shi2020improving,lee2025offer,henderson2022should,basik2018fair}.
These two protocols also induce fundamentally different objective functions over notification sets. Under BA, adding more drivers could only improve the score of the selected driver in expectation and always worsen the match time for this particular ride. Under FA, the valuation can be \emph{non-monotone}: notifying an additional low-value but high-acceptance driver can reduce the expected score by increasing the probability that a higher-value driver loses in the contention resolution stage, but it improves the match time for this particular ride. 

Our research collaboration with Lyft centers on formalizing the non-exclusive dispatch optimization problem under FA and BA, and on quantifying the advantages and limitations of NED relative to traditional exclusive dispatch. Within a single dispatch cycle, relaxing exclusivity expands the feasible set---and thus can only improve the platform's best achievable objective---at the cost of increased computational complexity.\footnote{The focus of this work is largely on studying NED in a dynamic marketplace using simulations and a stylized macro-model. In our companion paper~\ifblind\cite{LyftI2026companion_Blind}\else\cite{LyftI2026companion_NonBlind}\fi, we instead focus on the algorithmic problem of single-cycle optimization and provide a theoretical study of approximation algorithms for this combinatorial optimization problem.} Over time, however, broadcasting can also ``reserve'' multiple drivers for the same request, thinning effective supply and altering future market thickness. We therefore study---both empirically and theoretically---the long-run performance of NED, as well as the resulting speed--quality tension in ride-hailing dispatch under different design choices. These considerations motivate the following research questions.
\vspace{-5mm}
\begin{displayquote}
\emph{
\begin{enumerate}[label=(\roman*), align=left, labelwidth=1em,  
  labelsep=0.5em,         
  leftmargin=!,         
  itemsep=0.2em]
    \item Can non-exclusive dispatch fundamentally improve platform's key fulfillment metrics (match time, quality, and throughput) compared to the traditional exclusive dispatch?
\item If so, how to compare FA and BA in terms of these fulfillment metrics? Does the quality-speed trade-off between FA and BA contention resolution persist in equilibrium, or does one strategy strictly dominate the other 
over time?~\footnote{As we elaborate later, we analyze the system under the assumption that drivers' acceptance behavior remains unchanged under the new notification system. We postpone a combined analysis of fulfillment metrics and behavioral responses to future work.}
\end{enumerate}
}\end{displayquote}

\subsection{Our Contributions}
We provide compelling answers to the above questions using a novel two-layer empirical-theoretical framework. We first build a high-fidelity discrete-event simulator for NED calibrated to Lyft data (\Cref{sec:simulation}). We then develop a stylized macro model that endogenizes market thickness and provides theoretical validation (intuition) for the empirical patterns in our simulations (\Cref{sec:MMM}). Concretely, we make three contributions:


\smallskip
\xhdr{I. Unified framework for non-exclusive dispatch (\Cref{sec:prelim}):}
We formalize non-exclusive dispatch with stochastic driver participation and decompose NED into (i) a notification packing combinatorial optimization problem and (ii) a contention resolution protocol (under FA, BA, and the interpolating $k$-Accept family). This decomposition clarifies how operational rules induce different per-ride welfare objectives over notification sets and separates single-cycle optimization from long-run marketplace effects, allowing us to analyze dispatch protocols as independent decision layers. 


\smallskip
\xhdr{II. Simulation study of long-run marketplace effects using Lyft data (\Cref{sec:simulation}):}
We build a discrete-event simulation environment using proprietary Lyft data (including rider- and driver- dependent events' timestamps and dispatch data) and use it to benchmark a wide range of non-exclusive dispatch protocols. We rigorously evaluate multiple notification packing strategies (\textsc{Opt}, \textsc{Approx}, \textsc{Greedy}, \textsc{Rejection-aware}, and \textsc{Ed+}; see Appendix~\ref{sec:Algs}) under FA/BA/$k$-Accept contention rules by testing their performance based on key fulfillment metrics. The simulator models asynchronous accept/reject decisions, notification timeouts, and rider/driver abandonment, and it addresses key challenges such as counterfactual scoring and computational scalability (\Cref{sec:Data}). It also provides the basic means of extending the single-cycle performance evaluation of any NED notification algorithm to its long-run performance, which we exploit next.


\smallskip
\xhdr{III. Stylized marketplace macro model for theoretical analysis (\Cref{sec:MMM}):} 
To understand the fundamental forces driving system performance in the long-run and to isolate which effects persist once market thickness is endogenized, we introduce the 
\emph{Marketplace Macro Model (MMM)} in \Cref{sec:MMM}. The MMM is a large-market fluid model represented as a continuous-time Markov chain (CTMC). This model provides a stylized analytical framework that captures the essence of the feedback loop between dispatch efficiency and market thickness. By solving for the steady-state equilibrium of the MMM under both FA and BA, we explain the relationship between market density and matching efficiency, providing theoretical grounding and intuition for our simulation results. Specifically, the MMM yields steady-state match probabilities and conditional match times through CTMC calculations (\Cref{prop:MatchTime}) and links these quantities associated with any NED algorithm (composed of a packing algorithm and a contention resolution protocol) to the equilibrium market thickness induced by that algorithm through an iterative fixed-point method. Using a simulation-based computational approach, we also demonstrate how this tool can be used to translate ``local'' single-cycle behavior of any NED algorithm into its performance based on ``global'' long-run metrics.

We conclude in Section~\ref{sec:conclusion} with a discussion of future directions. A more detailed review of related work appears in Appendix~\ref{app:related-work}.

\section{NED Notifications Algorithmic Framework: Preliminaries}
\label{sec:prelim}
This section formalizes the within-cycle model used throughout the paper. Building on the high-level description in \Cref{sec:intro}, we describe the cycle-based dispatch pipeline and highlight the components most relevant for non-exclusive notifications. Our goal is not to reproduce Lyft's full matching production system; instead, we isolate first-order features that directly affect match quality, driver responses, and rider match time.


\smallskip
\noindent\textbf{Cycle-based matching optimization at Lyft.}  
At a high level, the matching pipeline at Lyft, similar to other ride-hailing platforms such as Uber and DiDi, operates  in real-time to connect riders and drivers. Lyft runs dispatch in discrete \emph{cycles} of fixed length (e.g., every few seconds) within each geographic region (e.g., New York city). In each cycle, the platform selects which drivers to notify for which riders, sends notifications, and collects accept/reject decisions (or timeouts). Matches are finalized asynchronously later, once a driver accepts. This process repeats over discrete cycles, allowing the system to adapt to changing demand, supply, traffic conditions, and strategic behaviors.

During a cycle, the platform considers (i) the set of active rides on the demand side, denoted by $\riders$, and (ii) the set of available drivers on the supply side, denoted by $\drivers$. The riders in $\riders$ have requested a trip in the past but have not yet been successfully matched with any driver.
The drivers in $\drivers$ are idle (i.e., not en route to a destination)\footnote{One simplification in our model is that we consider only ``simple'' matches between new riders and idle drivers. In practice, platforms also use more complex options such as linking a driver who is about to finish a trip to a new rider or swapping assignments between notified drivers. We abstract away from these features in our current analysis.} and are not currently in the \emph{notified} state (i.e., they have no pending notifications). Given sets $\riders$ and $\drivers$, the platform decides on the notifications to be sent at the end of the cycle.

Once a driver is notified, the notification appears in the driver’s app. Importantly, in the current system, each driver receives \emph{at most one} notification at a time. Under the baseline exclusive dispatch (ED) system, each ride request is also offered exclusively to at most one driver per cycle. A notified driver may accept or reject; if the driver does not respond within a fixed time window (e.g., ten or twenty seconds), the notification times out. While a driver is in the notified state, they are temporarily unavailable for future cycles; they return to the available pool once the notification is rejected, timed out, or withdrawn (the latter option is a feature specific to NED, as we will discuss later).

To evaluate the quality of a potential match, Lyft uses a scoring system that assigns a non-negative score $\revcolor{w_{r,d}}\in \mathbb{R}_{\geq 0}$ to each feasible rider–driver pair $(r,d)\in \riders\times\drivers$ within a cycle. These scores depend on factors such as predicted pickup time, trip duration, pay, and driver utilization, as well as other ride-specific features. At a high level, the goal of \emph{single-cycle matching optimization} is to decide which drivers to notify for which ride requests in that cycle so as to (myopically) maximize the total score of the eventual matches generated in that cycle. We additionally study rider and driver based metrics, including \emph{match time}, defined as the time from rider's request to match, and \emph{driver idle time}, defined as the time from when a driver becomes available to when they are assigned a ride.

\smallskip
\noindent\textbf{Driver rejections.}
A key behavioral feature of this system is \emph{driver rejection}: a notified driver may reject (or ignore) a ride request. Rejections are operationally costly. On the supply side, while a driver is in the notified state, they are unavailable for other matches; on the demand side, a rejection delays service for the rider (i.e., increases match time) and typically triggers additional notifications in later cycles. We model this behavior by assuming that each driver $d$ independently accepts a notified ride  \revcolor{$r$} with probability $\revcolor{p_{r,d}}$ and draws a response time $\tau$ independently from a common distribution $\Lambda$. While our framework allows for rider--driver-specific acceptance probabilities $p_{rd}$, we abstract away from this dependence in much of the analysis for simplicity. This distinction motivates two regimes: \emph{heterogeneity-aware} matching, which explicitly accounts for heterogeneity in the estimated acceptance probabilities in the marketplace (i.e., uses heterogeneous estimates), and \emph{homogeneous-acceptance} matching, which assumes a common acceptance probability estimate $p$ for all drivers (e.g., for fairness or simplicity).\footnote{Typically, probability $p$ is calibrated to represent the estimated marketplace average.} We study both regimes and examine how driver rejection interacts with matching decisions under each assumption.

\smallskip
\noindent\textbf{Exclusive vs.\ non-exclusive notifications.}
Earlier, we described the standard ED notifications, under which each ride request is offered to at most one driver at a time. In contrast, under \emph{non-exclusive dispatch} (NED), the platform may notify multiple drivers about the same request within each cycle (without committing the ride to any of them upfront), while still ensuring that each driver receives at most one notification. Broadcasting hedges against driver rejection by eliciting responses in parallel, but temporarily locks more drivers in the notified state and requires an additional contention-resolution step. Optimizing NED within a cycle is therefore naturally two-stage, and any procedure for this two-stage optimization problem should consist of two algorithmic components:

\begin{enumerate}[label=(\roman*)]
    \item \textbf{\emph{Packing algorithm}}: in the first stage, pick a many-to-one assignment between drivers $\drivers$ and riders $\riders$ in that cycle, which is equivalent to selecting a collection of disjoint notification sets $\{S_r\}_{r\in \riders}$ with  $S_r\subseteq \drivers$ and $\forall r,r'\in\riders: S_r\cap S_{r'}=\emptyset$.
    \item  \emph{\textbf{Contention resolution protocol}}: in the second stage, once notifications are sent, resolve the potential contention among the accepting drivers for each ride, i.e., decide which accepting driver---if any---should be assigned to (pick up) the rider corresponding to that request.
\end{enumerate}
We next examine these two stages and their algorithmic implications, which provide the building blocks for the simulation and macro-model analyses in \Cref{sec:simulation} and \Cref{sec:MMM}.

\subsection{Packing Algorithm}
\label{sec:Packing}
In each cycle, the platform observes a bipartite graph whose edges connect unmatched (or active) ride requests to available drivers. Each feasible rider--driver pair $(r,d)$ has a weight $\revcolor{w_{r,d}}$ capturing the match score, and an acceptance probability $\revcolor{p_{r,d}}$. A packing algorithm partitions drivers into disjoint notification sets and assigns one set to each ride. This static decision can be framed as a welfare maximization problem: each ride $r$ has an associated set function $F_r(\cdot)$ (think of it as a valuation function), induced by the specific contention-resolution protocol used in the second stage (\Cref{sec:single-cycle-contention-resolution}), and the goal is to choose disjoint sets $\{S_r\}_{r\in\rides}$ to maximize the total \emph{welfare} $\sum_{r\in\rides} F_r(S_r)$.

Formally, we consider the following constrained welfare maximization problem:
\begin{align} \tag{\textsc{Welfare-Max}} \label{prob:packing}
    \max_{\{S_r\}_{r \in \rides}} & \sum_{r\in \rides} F_r(S_r) \\
    \text{s.t.} \quad 
    & S_r \cap S_{r'} = \emptyset, && \forall r,r'\in \rides, r \neq r' \nonumber \\
    & |S_r| \leq U, && \forall r\in \rides \nonumber \\
    & F_r(S_r) - F_r(S_r\setminus \{d\}) \geq \theta \cdot \revcolor{p_{r,d}}, && \forall r\in \rides, d\in S_r. \nonumber
\end{align}
The objective function maximizes the aggregate expected welfare across all active requests in the current cycle. The first constraint guarantees that the notification sets are disjoint, ensuring that no driver receives more than one notification simultaneously. The second constraint limits the number of drivers notified for any single request to a system-defined maximum $U\in\mathbb{N}$, preventing excessive traffic (i.e., notification overload). Finally, the third constraint enforces the marginal efficiency condition: a driver $d$ is only included in the set $S_r$ if their contribution to the welfare function $F_r$ exceeds the weighted opportunity cost $\theta \cdot \revcolor{p_{r,d}}$. Here, $\theta$ is a  threshold parameter to control the future value of leaving a driver un-notified in the current cycle. This notification process not only avoids overwhelming drivers with redundant rides but also strategically manages the availability of drivers across time, especially in under-supplied conditions.\footnote{In other words, introducing parameters $U$ and $\theta$ helps with making the matching decisions to some extent non-myopic. As we show later in our analysis, this feature has benefits in terms of long-term performance.}

A central challenge in this phase is balancing short-term gain with long-term efficiency. While it might seem wasteful not to notify a driver who could plausibly accept a ride, withholding notifications can also preserve that driver’s availability for higher-value matches in future cycles. More precisely, although a solution to \eqref{prob:packing} with $U=\infty,~\theta = 0$ is optimal for a single-cycle, to account for the long-term impact of each decision over multiple cycles, we can tune parameters $U$ and $\theta$ to better reflect existing opportunity costs and future consequences (see our simulations in \Cref{sec:simulation} regarding varying these parameters).

Although this optimization problem is computationally hard in its general form\footnote{This hardness result is formally established in our companion paper ~\ifblind\cite{LyftI2026companion_Blind}\else\cite{LyftI2026companion_NonBlind}\fi, under both FA and BA.}, practical instances remain tractable due to the inherent sparsity of the underlying bipartite graph, which is enforced by type filters and maximum dispatch radii. This will be discussed more in \Cref{sec:simulation}. 

\subsection{Contention-Resolution Protocols}
\label{sec:single-cycle-contention-resolution}
While contention resolution can theoretically function as a complex, real-time adaptive process, we focus our analysis on two canonical non-adaptive rules. These protocols are selected for their simplicity and interpretability, yet they effectively capture the fundamental trade-off between decision speed and match quality. Under \emph{first-accept (FA)}, the ride is immediately assigned to the first driver who accepts the notification. Under \emph{best-accept (BA)}, the platform waits until all notified drivers have either responded or timed out, and then assigns the ride to the accepting driver with the highest score.
We derive the value functions for both (and numerically evaluate protocols interpolating between FA and BA, the details of which are in \Cref{sec:simulation}).\footnote{Note that lower match quality in a cycle does not necessarily lead to overall lower match quality in the long-run, as the market condition, i.e., how thick the market is, can also affect the quality of the match---which itself depends on how much supply is locked by the notification algorithm. We study this phenomenon in \Cref{sec:simulation} and \Cref{sec:MMM}.} 

\textbf{First Accept.} This approach prioritizes speed and is simple to implement, requiring minimal coordination. It naturally incentivizes quick responses from drivers, helping reduce matching delays. Additionally, it limits the number of drivers kept in notification mode. However, this strategy leads to potentially lower match quality in that cycle, and does not always yield the most efficient or optimal match; for instance, a slower-responding driver might be closer or better suited to fulfill the ride. The expected score for each rider $r$ by sending the notification to set $S$, in the case of using FA, is 
\begin{align}\label{eq:FA-value}\tag{FA}
    F_r(S) = \displaystyle\sum_{T \subseteq S} \left(\prod_{d \in S \setminus T} (1-\revcolor{p_{r,d}})\right)\left(\prod_{d \in T} \revcolor{p_{r,d}} \right)\frac{\sum_{d \in T} \revcolor{w_{r,d}}}{|T|}
\end{align}
The summation aggregates over all possible subsets $T \subseteq S$ representing the drivers willing to accept the request. The product terms define the probability that exactly the drivers in $T$ are willing to accept, assuming independent decisions. Conditional on the set $T$, the term $\frac{\sum_{d \in T} \revcolor{w_{r,d}}}{|T|}$ reflects the expected match value. This derivation relies on the assumption that response times among willing drivers are independent and identically distributed; consequently, under the FA mechanism, each driver in $T$ has an equal probability $1/|T|$ of responding first and securing the ride.

\smallskip
\textbf{Best Accept.} The BA protocol introduces a short waiting window to collect all driver responses before assigning the ride to the most suitable one based on a scoring function such as proximity. While this method delays immediate assignment, it can yield higher-quality matches and improve overall system efficiency. On the other hand, it may temporarily tie up multiple drivers (who accepted) without ultimately assigning them the ride, which can lead to dissatisfaction for those drivers. Additionally, since speed is not directly rewarded, drivers may be less motivated to respond promptly. The expected score for each rider $r$ by sending the notification to set $S$, in the case of using BA, is 
\begin{align*}\label{eq:BA-value}\tag{BA}
    F_r(S) = \displaystyle\sum_{T \subseteq S} \left(\prod_{d \in S \setminus T} (1-\revcolor{p_{r,d}})\right)\left(\prod_{d \in T} \revcolor{p_{r,d}} \right)\max_{d\in T}\{\revcolor{w_{r,d}}\}
\end{align*}
Analogous to the FA formulation, the outer summation traverses all potential subsets $T \subseteq S$ of drivers who affirmatively respond within the waiting window. The product terms capture the likelihood of observing exactly the set $T$ of willing drivers, assuming independent acceptance probabilities. The crucial distinction lies in the valuation term $\max_{d\in T}\{\revcolor{w_{r,d}}\}$. Unlike FA, where the match is effectively determined by reaction time, the BA strategy aggregates all responses and assigns the ride to the optimal driver within the accepting set $T$. Consequently, the realized utility for a given realization $T$ is the maximum weight among all willing drivers rather than the average.

Finally, we highlight a key structural distinction between the two strategies: unlike \eqref{eq:FA-value}, the \eqref{eq:BA-value} objective function is an increasing and submodular set function.\footnote{See our companion paper for examples showing FA valuations are neither monotone nor submodular. Also, they are not even ``score-ordered,'' which makes it difficult to design algorithms for this class~\ifblind\citep{LyftI2026companion_Blind}\else\citep{LyftI2026companion_NonBlind}\fi.} This property is particularly advantageous for the first-stage packing problem, as submodularity allows us to leverage the efficient approximation algorithms known in the literature. In fact, the unconstrained formulation of \eqref{prob:packing} corresponds to the well-known Submodular Welfare Maximization~\citep{vondrak2008optimal}. In this setting, efficient polynomial-time algorithms are guaranteed to yield a $(1 - 1/e)$-approximation of the optimal solution. Furthermore, simple greedy algorithms (Algorithm~\ref{alg:greedy} in Appendix~\ref{sec:Algs}) achieve a $1/2$-approximation in the general unconstrained case.

\smallskip
\textbf{\textit{k}-Accept.} We also consider a generalized contention resolution protocol, denoted as $k$-Accept, which serves as a hybrid between the pure speed of FA and the quality maximization of BA. Under this policy, the platform broadcasts notifications to the set $S$ and finalizes a match as soon as it receives at least $k$ \revcolor{acceptances}, where $1 \leq k \leq U$ is a tunable parameter. From this subset of accepted drivers, the system assigns the rider to the candidate with the highest utility. If fewer than $k$ drivers accept, the platform selects the highest-weight acceptor once all responses have arrived or the response window has closed; if no driver accepts, the rider remains unmatched. This protocol offers a flexible control lever: low values of $k$ minimize driver holding time, while higher values increase the likelihood of finding a high-quality match. Crucially, this policy recovers the previous strategies at its boundaries: setting $k=1$ reduces to \eqref{eq:FA-value}, while setting $k=U$ (or $k=|S|$) forces the system to consider the entire set $T$, recovering the maximum weight objective of \eqref{eq:BA-value}.

\section{A Large-Scale Discrete-event Simulation with Lyft Data}
\label{sec:simulation}
As described in the introduction, the main research goal in our collaboration with Lyft is to empirically and theoretically quantify the \emph{long-run performance} of NED notifications in the Lyft marketplace. In this section, we focus on the empirical component: we evaluate the impact of NED on fulfillment metrics in a long window of time (\revcolor{twenty five} minutes) under the counterfactual assumption that drivers’ acceptance behavior does not change in response to the notification mechanism (i.e., we ignore temporary behavioral responses). 
Among other things, this empirical study provides a rigorous way to assess the potential impact of NED marketplace intervention before running an online experiment, which is typically costly and operationally risky.\footnote{The focus of this section is on simulations; we defer the stylized equilibrium analysis to \Cref{sec:MMM}.}

At the core of our empirical study, we build a large-scale discrete-event simulation environment using a proprietary Lyft dataset. We use this simulator to compare strategies that differ in both their packing algorithms and contention-resolution rules. Throughout, we distinguish between \RNcolor{heterogeneity-aware policies, which leverage heterogeneous acceptance probability estimates $\revcolor{p_{r,d}}$, and homogeneous-acceptance policies, which assume a uniform acceptance probability estimate $p$ across the population.}

We aim to empirically answer the following questions:
\begin{enumerate}[label=(\arabic*), align=left, labelwidth=0.2em,  
  labelsep=0.5em,         
  leftmargin=3em,         
  itemsep=0.1em]
    \item \emph{Considering key fulfillment metrics---match quality, rider match time, and throughput---do NED notifications outperform the status-quo exclusive dispatch?}
    \smallskip
    \item \emph{Is there an inherent trade-off between speed and quality in NED notification systems? How do design choices (in particular FA vs.\ BA) impact this trade-off in the long run?}
\end{enumerate}
The central findings of our simulations (later rationalized by the MMM in \Cref{sec:MMM}) are as follows:
\begin{enumerate}[label=(\roman*), align=left, labelwidth=1em,  
  labelsep=0.5em,         
  leftmargin=2em,         
  itemsep=0.2em]
    \item \textit{NED dominates ED on fulfillment metrics:} Across the tested configurations, NED simultaneously reduces match time (which reduces rider waiting and reneging) and improves the number of finalized matches, while also improving average realized match quality, relative to ED. Therefore, holding drivers' behavior fixed, NED dominates ED in all key fulfillment metrics.

    \item \textit{A speed--quality trade-off persists between FA and BA:} FA leads to shorter match times, hence faster and often more finalized matches (higher throughput). In contrast, BA produces higher \emph{average} match quality, but with longer match times (and hence higher rider waiting/reneging).
    
    \item \textit{Throughput can dominate average quality.} Even when BA has higher per-match quality, FA might lead to higher \emph{total} welfare (i.e., total matching score). This reversal is driven by system throughput, that is, completing more matches in the same time window. The cumulative gain from these additional matches could then outweigh the marginal loss in individual match quality. 
    \item \textit{Packing choices affect long-run availability.} Allowing minor sub-optimality in the algorithm for immediate packing problem---e.g., being conservative via opportunity costs/thresholding,  or using simple heuristics instead of optimal-per-cycle packing---can improve long-run efficiency by avoiding excessive locking of high-value drivers.
\end{enumerate}
We defer the discussion of managerial insights from our results and ideas to improve the system to \Cref{apx:simulation-discussion}.

\subsection{Simulation Setup: Benchmarks, Data \& Assumptions}
\label{sec:simulation-setup}
\subsubsection{NED packing algorithms.} Throughout our simulations, we evaluate the performance of several distinct algorithmic strategies (inspired by practice at Lyft). Fixing the contention resolution protocol (FA or BA), we employ the following algorithms: (i)~\textsc{Opt}, which is the single-cycle optimal algorithm and is implemented by solving an integer program, (ii) \textsc{Greedy} (Algorithm~\ref{alg:greedy}), which iterates through the available drivers and tentatively assigns each to the ride that yields the highest marginal increase in the expected welfare, (iii) \textsc{Rejection-aware} (Algorithm~\ref{alg:rej_aware}), which is a heuristic designed to approximate practical dispatch logic used in industry, (iv) \textsc{Ed+} (Algorithm~\ref{alg:enhanced_nd}), which is a hybrid strategy that improves upon the standard ED baseline by adding a second driver from remaining unassigned pool if this increases the expected welfare, and (v) \revcolor{\textsc{Approx} \ifblind\citep{LyftI2026companion_Blind}\else\citep{LyftI2026companion_NonBlind}\fi, which solves the LP relaxation and applies a derandomized rounding procedure to construct the notification sets. In \ifblind\cite{LyftI2026companion_Blind}\else\cite{LyftI2026companion_NonBlind}\fi, it is shown that this algorithm achieves the theoretical $(1-1/e)$ approximation guarantee in the untrimmed setting (i.e., with $U=\infty$ and $\theta=0$; see Appendix~\ref{sec:Algs})}. In particular, we consider parametric versions of these algorithms with two parameters $(U,\theta)$, where $U$ denotes the maximum size of any notification \revcolor{set}, and $\theta$ denotes a threshold quantifying the estimated opportunity cost of assigning a driver \revcolor{(i.e., the threshold in the marginal efficiency constraint of \eqref{prob:packing}, which each algorithm enforces in its own way; \textsc{Rejection-aware} uses only $U$)}. See Appendix~\ref{sec:Algs} for further details around these algorithms.

\smallskip
\subsubsection{Lyft dataset: details \& challenges.} \label{sec:Data} Our experimental evaluation utilizes a dataset of ride requests and driver trajectories from a US metropolitan area, captured during an afternoon window (4:00 PM -- 4:25 PM) on a weekday. To adapt these historical data for our simulation environment, we address two primary challenges regarding counterfactual scoring and computational scalability. In short, we show how to use position data to obtain a proxy score for each pair of riders and drivers (even those that are not present in our matching data). We also show how to use techniques such as graph decomposition and pruning to reduce the computational complexity of the NED optimal packing problem. 

\xhdr{Proxy scoring and compatibility.}
A fundamental limitation of using historical logs for simulation is the observational bias inherent in the data: we record only the outcomes of realized decisions, while the responses to alternative (counterfactual) offers remain unobserved. Historical datasets record compatibility scores only for rider-driver pairs that overlapped in time during the actual operations. However, our proposed algorithms may make different decisions than the production system—for instance, choosing to keep a driver idle rather than dispatching them immediately. In such a case, the simulation might retain a driver $d$ until a later rider $r$ arrives, creating a potential match $(r, d)$ that never occurred in reality because $d$ was already busy. Since the historical logs contain no score for this non-existent interaction, we cannot rely on the proprietary scoring.

To overcome this, we reconstruct the simulation environment using the raw spatial coordinates of all agents at their arrival times. We define a proxy compatibility score based on the inverse of the Euclidean distance:\footnote{This stylized choice of a proxy score based on geographical location is due to the fact that we cannot use Lyft's original scoring system in a simulation with counterfactual pairs due to certain technical limitations beyond the scope of this work}.
\begin{align}\label{scoreDist}
    \revcolor{w_{r,d}} = \frac{1}{1 + \text{dist}(r, d)}.
\end{align}
This allows us to compute a consistent weight for \emph{any} feasible pair within a standard dispatch radius, creating a dense and realistic testbed for comparing our algorithms.

\smallskip
\xhdr{Graph decomposition and pruning.}
To ensure the optimal packing algorithm OPT (\Cref{sec:optALG}) remains tractable, i.e. solvable within seconds, we exploit the structural properties of the compatibility graph to significantly reduce the problem dimension.

\paragraph{Connected components:}
First, we leverage the spatial sparsity of the graph. Because matches are constrained by a maximum dispatch radius and spatial filters, the global compatibility graph naturally fractures into multiple disjoint connected components. For example, the matching dynamics in one neighborhood are often completely independent of those in a distant borough. We identify these components and solve the optimization problem for each independently, allowing for massive parallelization.

\paragraph{Lossless pruning:}
Second, we implement a pruning procedure for degree-one nodes (exclusive drivers). For a rider $r$ with a set of exclusive drivers $D_r$, any optimal notification set of size $\ell \leq U$ must necessarily comprise the highest-utility drivers within each acceptance probability class. This dominance property allows us to immediately discard any driver who does not rank within the top $U$ of their specific class. Consequently, the search space is drastically reduced: rather than evaluating all $\binom{|D_r|}{\ell}$ combinatorial subsets, we need only consider the top-ranked candidates (across the possible probability classes/driver types, explained next in \Cref{apx:modeling-agent}). This simplification significantly lowers the computational complexity of the integer program.


\smallskip
\subsubsection{Modeling assumptions.}\label{apx:modeling-agent}
We adopt a few simplifying assumptions regarding riders and drivers behavior. Overall, our simulation assumes heterogeneous driver acceptance behavior (four driver types), no cancellations after a match is finalized, and stochastic abandonment for unmatched agents. Together, the following assumptions keep the simulator tractable while capturing essential market dynamics and frictions, i.e., driver selectiveness and the time cost of waiting. 


\begin{enumerate}[label=(\roman*), align=left, labelwidth=1em,  
  labelsep=1em,         
  leftmargin=!,         
  itemsep=0.0em]
    \item \textit{Driver acceptance heterogeneity:} \revcolor{To capture the variance in driver preferences, arising from unobserved factors such as earnings targets, we assume drivers are not uniform in their willingness to accept requests. Each driver is assigned a base acceptance probability ${\rho_d} \in \{0.1, 0.3, 0.6, 0.9\}$ with population proportions $\{0.1, 0.3, 0.3, 0.3\}$, calibrated to Lyft's driver data, and each ride draws a quality multiplier $\gamma_r \in \{0.5, 1, 1.5\}$ with equal probability. Driver $d$ accepts a notification for ride $r$ with probability $\revcolor{p_{r,d}} = \min\{\revcolor{\rho_d\, \gamma_r},\ 0.95\}$.}

    \item \textit{Driver response time:} To simulate the operational latency inherent in driver interaction with the app, we introduce a stochastic decision window. Upon receiving a notification, a driver is allowed a maximum of $7$ simulation cycles to accept or reject the offer. We model the actual response time as a uniform random variable drawn from the set $\{1, \dots, 7\}$ cycles. This mechanism ensures that matches are not instantaneous, reflecting the real-world friction of driver availability and safety constraints.
    
    \item \textit{Post-match commitment: } We assume perfect reliability for realized matches; that is, once a driver and rider are successfully paired by the platform, neither party cancels the trip. This assumption allows us to isolate the efficiency of the matching algorithm from the noise of operational friction, e.g., cancellations due to traffic or personal preference.
\item \textit{Driver/rider patience and abandonment:}
We model the patience of unmatched agents using a probabilistic decay. In each simulation cycle (set to 3 seconds), any available agent who remains unmatched faces a risk of abandoning the platform. Specifically, we assume that an idle driver exits with a probability of $\tilde{\eta}_I=0.001$ per cycle (and a notified driver, who has a pending notification, does not abandon the system while in this state, i.e., $\tilde{\eta}_N=0$), while an unmatched rider---who is usually more time-sensitive---exits with a higher probability of $\eta=0.01$ per cycle. This mechanism prevents the artificial accumulation of inventory and reflects the real-world opportunity costs faced by market participants.
\end{enumerate}

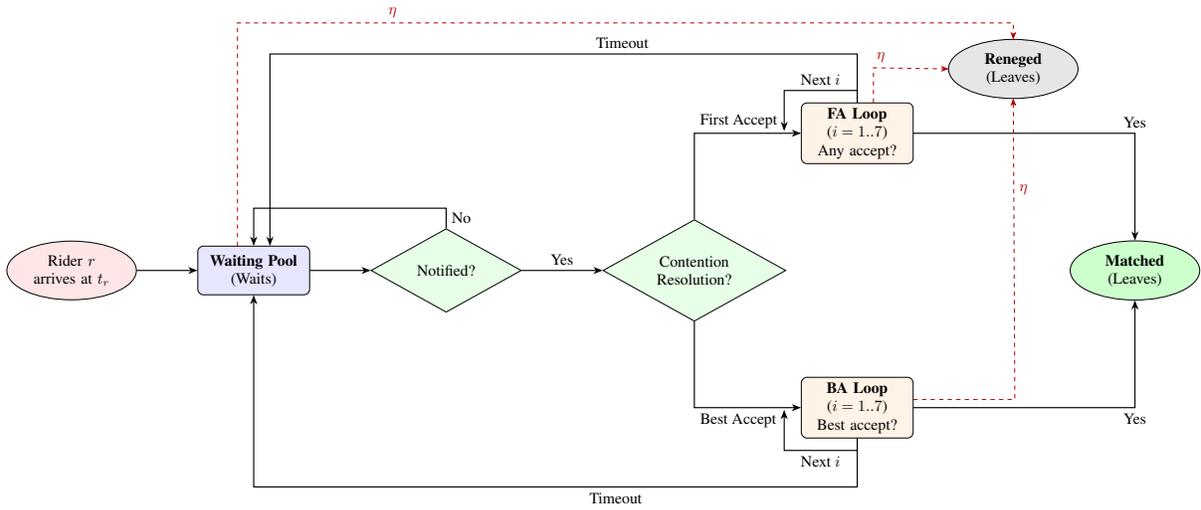
\begin{figure}[htb]
    \centering
    \begin{center}
\resizebox{0.95\textwidth}{!}{
    \begin{tikzpicture}[
        node distance=1.5cm and 1.5cm,
        startstop/.style={ellipse, draw, fill=red!10, text width=2cm, align=center, minimum height=1cm},
        process/.style={rectangle, draw, fill=blue!10, text width=2.5cm, align=center, minimum height=1.2cm, rounded corners},
        decision/.style={diamond, draw, fill=green!10, text width=2.5cm, align=center, aspect=1.8},
        loopnode/.style={rectangle, draw, fill=orange!10, text width=2.5cm, align=center, rounded corners},
        arrow/.style={thick, ->, >=Stealth},
        reneg/.style={dashed, ->, >=Stealth, red!70!black}
        ]

        
        \node (start) [startstop] {Rider $r$ arrives at $t_r$};
        \node (pool) [process, right=of start] {\textbf{Waiting Pool} \\ (Waits)};
        \node (notified) [decision, right=of pool] {Notified?};
        \node (protocol) [decision, right=of notified, xshift=0.5cm] {Contention Resolution?};
        
        \node (fa_loop) [loopnode, above right=of protocol, yshift=0.5cm] {\textbf{FA Loop} \\ ($i=1..7$) \\ Any accept?};
        \node (ba_loop) [loopnode, below right=of protocol, yshift=-0.5cm] {\textbf{BA Loop} \\ ($i=1..7$) \\ Best accept?};
        
        \node (matched) [startstop, fill=green!20, right=of protocol, xshift=5.5cm] {\textbf{Matched} \\ (Leaves)};
        
        \node (leaves_reneg) [startstop, fill=gray!20, above=of matched, xshift=-3cm , yshift=2cm] {\textbf{Reneged} \\ (Leaves)};

        
        \draw [arrow] (start) -- (pool);
        \draw [arrow] (pool) -- (notified);
        \draw [arrow] (notified) -- node[above] {Yes} (protocol);
        
        \draw [arrow] (notified.north) -- node[right] {No} ++(0, 0.5) -| (pool.north);
        
        \draw [arrow] (protocol.north) |- node[above right] {First Accept} (fa_loop.west);
        \draw [arrow] (protocol.south) |- node[below right] {Best Accept} (ba_loop.west);
        
        \draw [arrow] (fa_loop.east) -| node[above] {Yes} (matched.north);
        
        \draw [arrow] (fa_loop.north) -- ++(0,0.3) -- node[above] {Next $i$} ++(-1.8,0) -- ++(0,-1) coordinate (corner);
        
        \draw [arrow] (fa_loop.north) -- ++(0, 1.2) -| node[above, pos=0.2] {Timeout} ($(pool.north) + (0.4, 0)$);
        
        \draw [arrow] (ba_loop.east) -| node[below] {Yes} (matched.south);
        \draw [arrow] (ba_loop.south) -- ++(0,-0.3) -- node[below] {Next $i$} ++(-1.8,0) -- ++(0,1) coordinate (corner);
        \draw [arrow] (ba_loop.south) -- ++(0, -1.2) -| node[below, pos=0.2] {Timeout} (pool.south);

        
        \draw [reneg] ($(pool.north) + (-0.4, 0)$) -- ++(0, 5.5) -| node[above, pos=0.1] {$\eta$} (leaves_reneg.north);
        
        \draw [reneg] ($(fa_loop.north) + (0.4, 0)$) -- ++(0, 0.86) -- node[above, pos=0.1] {$\eta$} (leaves_reneg.west);
        
        \draw [reneg] ($(ba_loop.east) + (0, 0.2)$) -- ++(2.5, 0) -- node[right, pos=0.7] {$\eta$} (leaves_reneg.south);

    \end{tikzpicture}
}
\end{center}
    \caption{Flowchart illustrating the simulation lifecycle from a rider's perspective. (Note: Under BA if the highest-ranked driver accepts prior to the 7th cycle, the match is finalized immediately.)}
    \label{fig:flowChart}
\end{figure}

Figure \ref{fig:flowChart} depicts the simulation logic from the viewpoint of a rider. Upon arrival at time $t_r$, the rider enters a waiting pool. Throughout the process, whether in the waiting pool or during a matching attempt, the rider may renege and permanently leave the system with probability $\eta$. Once a notification set is drawn from the idle driver pool, the process follows one of two contention resolution protocols. In the FA branch, the rider is matched immediately upon the first acceptance. In the BA branch, the system aggregates responses and assigns the highest-scoring driver. Both protocols operate within a maximum window of 7 cycles ($i=1\dots7$); if no match is finalized by the end of this window, the rider returns to the waiting pool. The driver life cycle follows a parallel logic, changing between the idle and notified states as they accept or reject notifications.

\subsection{Simulation Results}\label{sec:SimulationResults}
In this section, we present the comparative results of our simulations across the defined metrics. We sample 2000 problem instances with heterogeneous drivers, where, in each instance, the acceptance probability of each driver \revcolor{and the quality multiplier of each ride are} sampled independently from the meta distribution described in \Cref{apx:modeling-agent}. We then perform a Monte-Carlo simulation and measure certain aggregate metrics for each sampled instance. To rigorously evaluate the efficacy of the proposed algorithms, we focus our analysis on three primary aggregate performance indicators studied in ride-hailing fulfillment:
\begin{enumerate}[label=(\roman*), align=left, labelwidth=1em,  
  labelsep=1em,         
  leftmargin=!,         
  itemsep=0.2em]
    \item \textit{Average score:} The average score of all realized (i.e., successful) matches in our dispatching horizon, calculated using the distance-based proxy defined in equation~\eqref{scoreDist} (\Cref{sec:Data}).
    \item \textit{Average match time:} The average duration between a rider's arrival to the platform and their successful pairing with a driver who accepted the ride (and selected by NED), where the average is taken over all realized matches in our dispatching horizon. 
    
    \item \textit{Match count:} The total number of successful rider-driver pairings throughout the dispatching horizon determined by our dataset (around 25 minutes).
\end{enumerate}

\begin{figure}[h]
    \centering
    \includegraphics[width=\linewidth]{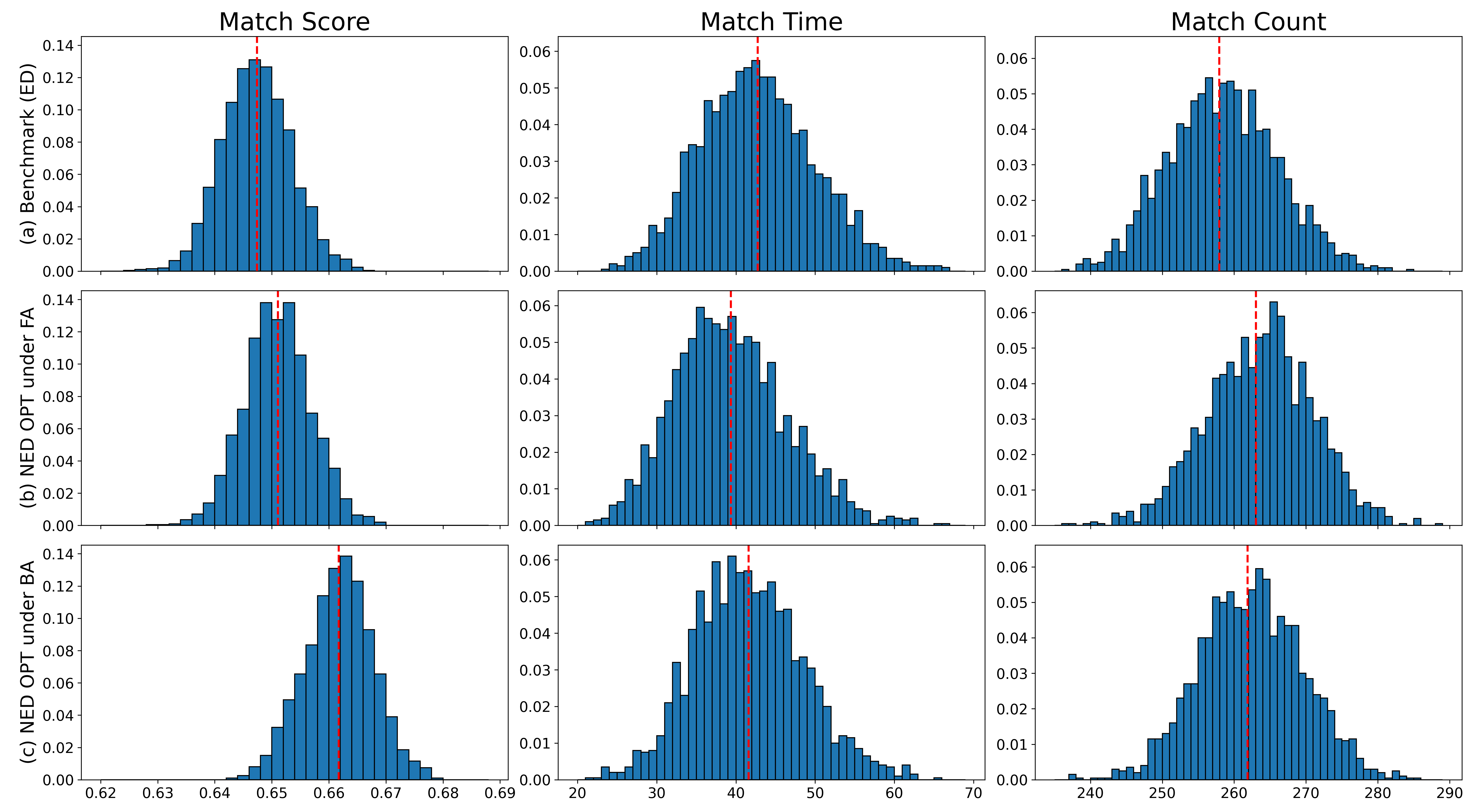}
    \caption{ED vs.\ NED with optimal packing ($U=3$, $\theta=0$) under FA and BA. Rows:
    (a) the ED benchmark, (b) NED OPT under FA, (c) NED OPT under BA.
    Columns: histograms, over 2000 sampled instances, of the average match score, the
    average rider match time, and the number of finalized matches (fraction of instances;
    dashed line = mean).}
    \label{fig:NE_Optimal}
\end{figure}

\smallskip
\xhdr{Baseline performance under exclusive dispatch.} \revcolor{We begin by establishing a baseline using the optimal exclusive-dispatch policy, which solves a maximum-weight matching\footnote{Throughout the remainder of the paper, ``maximum-weight matching'' under exclusive dispatch uses \textit{expected} edge weight $p_{r,d}w_{r,d}$: the match score multiplied by the driver's probability of accepting the request.} in each cycle. Row (a) of \Cref{fig:NE_Optimal} reports the \revcolor{histograms} of the three key metrics for this baseline across 2000 sampled problem instances.}

\revcolor{As shown in \Cref{fig:NE_Optimal}, the baseline achieves a long-run average score of a little less than 0.65, with an average match time of roughly 43 seconds and a total match count slightly below 260. This sets the reference standard against which we evaluate the NED algorithms.}

\smallskip
\xhdr{Performance of NED optimal packing algorithms.}
\revcolor{Next, we evaluate the performance of \revcolor{NED} using the same underlying optimization logic---that is, finding the optimal batch in a single cycle (NED OPT)---but relaxing the exclusivity constraint.  By allowing multiple notifications ($U=3$) with a standard threshold ($\theta=0$), we observe significant improvements.

Rows (b,c) of \Cref{fig:NE_Optimal} \revcolor{display} the performance histograms of the optimal packing algorithm under  FA and BA. Relative to the ED baseline, both NED OPT under FA and NED OPT under BA improve all three metrics in this setting. 
Two mechanisms contribute to this improvement. First, broadcasting reduces the time to obtain an acceptance, which decreases rider reneging and increases throughput. Second, as a managerial insight, relaxing the strict $U=1$ constraint avoids ``single-shot'' offers: high-value matches, such as a driver who is nearby and highly reliable in terms of acceptance, are not forfeited simply because another driver is marginally closer. Therefore, even if the highest-scoring candidate rejects, other plausible candidates can accept within the same notification window, potentially improving match probability and realized quality.

Based on the above findings, the two contention rules then separate along the intended dimensions. Under FA, the platform commits as soon as the first driver accepts, yielding the fastest matches (average match time around 39 seconds in our experiments). Under BA, the platform waits and then selects the highest-scoring accepting driver, producing the highest average score but with longer match times.}

\smallskip
\xhdr{Global performance landscape.} \revcolor{We synthesize the results by analyzing the trade-off between speed and quality across all tested algorithms/configurations. \Cref{fig:scatter} presents a scatterplot of average score (left) and total match count (right) versus average match time for various combinations of packing algorithms, contention protocols (FA/BA), notification size upper bounds ($U \in \{2,3,4,5\}$), and opportunity cost thresholds ($\theta \in \{0, \dots, 0.5\}$).
\begin{figure}[h]
    \centering
    \includegraphics[width=0.95\linewidth]{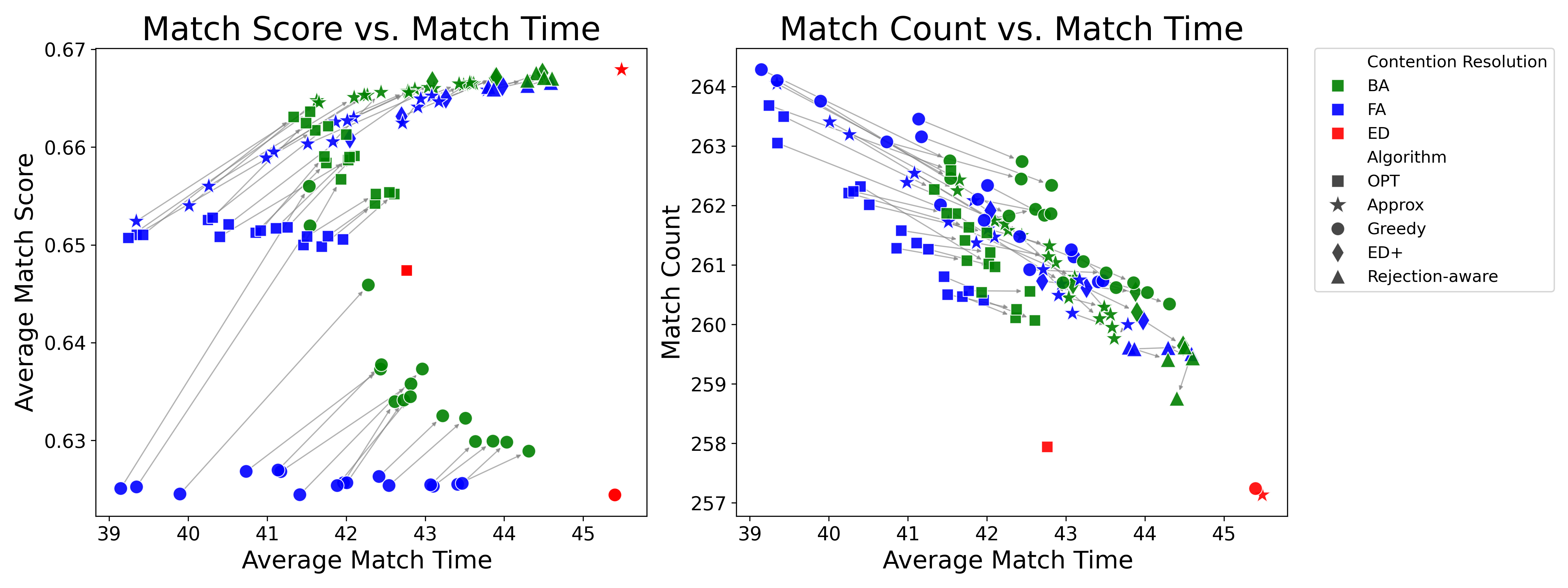}
    \caption{Performance of the algorithms across configurations. Left:
average match score vs.\ average match time. Right: match count vs.\ average match time. Each
point is one combination of algorithm (marker), contention resolution protocol (color),
$U\in\{2,\dots,5\}$ and $\theta\in\{0,\dots,0.5\}$. Gray arrows point
from the FA to the BA version of the same configuration. Red markers show the exclusive-dispatch versions ($U=1$) of \textsc{Opt}, \textsc{Approx}, and \textsc{Greedy} (same marker shapes); the red square (\textsc{Opt} with $U=1$) is the ED baseline of \Cref{fig:NE_Optimal}.}
    \label{fig:scatter}
    \vspace{-3mm}
\end{figure}

The results in \Cref{fig:scatter} provide a comprehensive comparison of the proposed algorithms, revealing distinct trade-offs between match time, total matches, and average score. 

First, we observe a clear distinction between the performance of the \textsc{Approx} algorithm (stars) and the statically optimal \textsc{Opt} algorithm (squares). Although \textsc{Opt} is designed to maximize welfare in a single static snapshot, it often underperforms \textsc{Approx} in the long run. This counterintuitive result highlights the difference between static and dynamic optimality: by aggressively exhausting the “best” available matches in the current cycle, \textsc{Opt} can deplete high-quality driver inventory that would have been better used to serve future demand. In contrast, \revcolor{the} \textsc{Approx} algorithm, by remaining slightly suboptimal in the short run, implicitly preserves system flexibility and achieves superior aggregate outcomes over time.

Second, the choice between the FA and BA contention-resolution protocols presents a clear speed--quality trade-off. The gray arrows in \Cref{fig:scatter} connect the FA and BA outcomes for the same algorithm and hyperparameter configuration, revealing a consistent shift from faster matching under FA to higher match quality under BA. As shown in the left panel, BA (green) achieves a higher average match score in all 56 configurations, improving the score by approximately $0.007$ on average, but increases average match time by approximately $1.05$ seconds. This delay translates into lower throughput: as shown in the right panel, FA (blue) causes more matches in 54 of the 56 configurations, with switching from FA to BA reducing the match count by approximately $0.6$ on average. These results reflect the underlying operational trade-off: FA resolves contention quickly and reduces rider reneging, whereas BA waits longer to secure a better-fitting driver.

Consequently, there is no single “dominant” strategy. The appropriate choice depends on the platform’s objective function. If the goal is to maximize market liquidity and throughput (match count), FA is preferable. If the platform instead prioritizes the quality of individual pairings (average score) and can tolerate modestly longer wait times, BA offers a distinct advantage. Note also that every tested NED configuration in \Cref{fig:scatter}, under both FA and BA, finalizes more matches than the exclusive-dispatch policies (shown in red), and that NED with optimal packing improves all three key metrics relative to the optimal ED baseline in every tested configuration, indicating that parallel notifications are the primary driver of the improvement; the FA/BA choice then governs how the gains are allocated between speed (FA) and match quality (BA).}

\smallskip
\xhdr{\RNcolor{Homogeneous-acceptance  policies}.} We extend our analysis to ``\RNcolor{homogeneous-acceptance}'' algorithms, \RNcolor{where the platform ignores heterogeneity in acceptance probability estimates} and instead utilizes a global statistical average for all acceptance probabilities. This formulation is particularly relevant for fairness considerations, as it ensures that dispatch decisions are structurally independent of a driver's specific history or behavioral profile. In short, (i) our results remain qualitatively similar in this setting, and (ii) we empirically demonstrate that the structural advantage of NED notifications outweighs the benefit of having perfect information in ED setting. See Appendix~\ref{apx:acceptance-unaware} for more details.

\begin{figure}[h]
    \centering
    \begin{subfigure}[t]{0.95\linewidth}
        \centering
        \includegraphics[width=\linewidth]{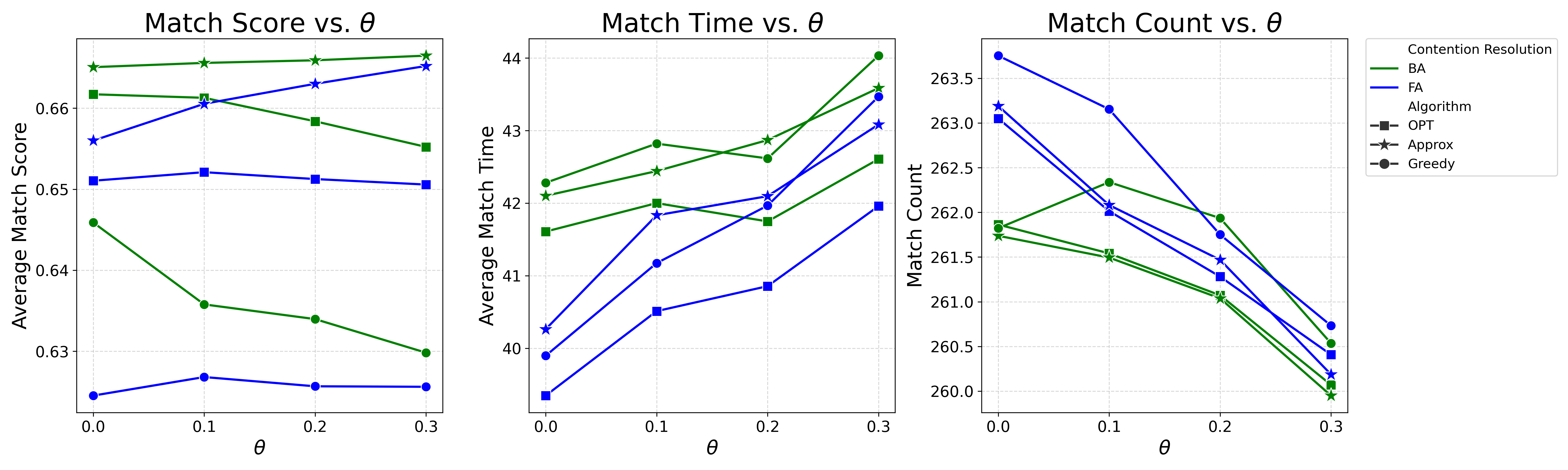}
        \caption{\footnotesize{Effect of varying $\theta$ (fixed $U=3$).}}
        \label{fig:theta_analysis}
    \end{subfigure}

    \vspace{-0.1em}

    \begin{subfigure}[t]{0.95\linewidth}
        \centering
        \includegraphics[width=\linewidth]{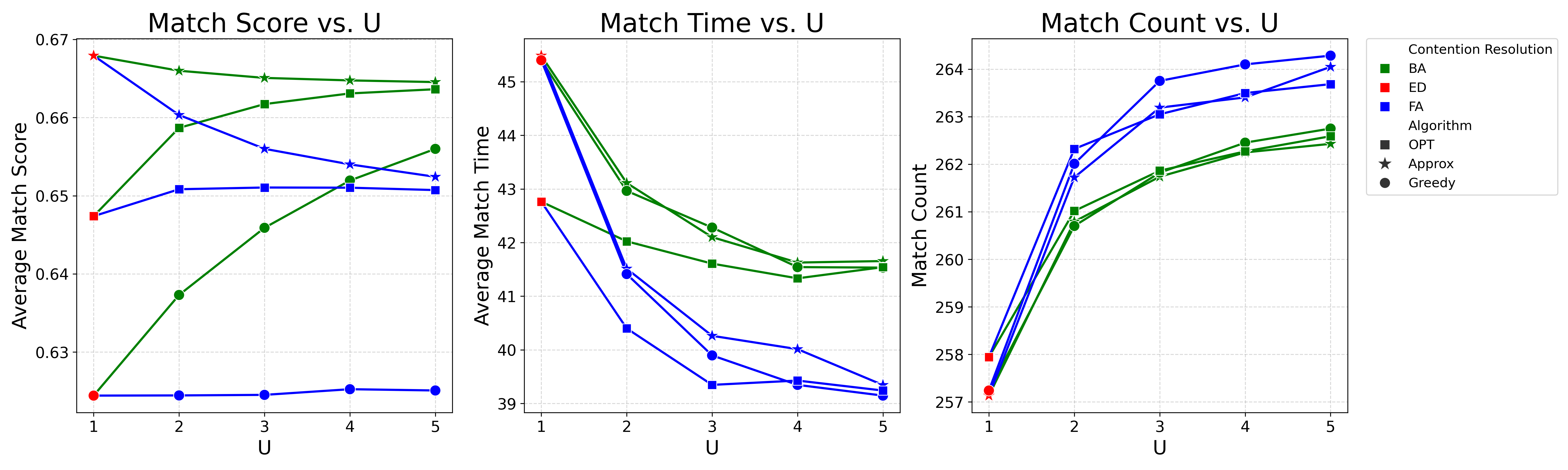}
        \caption{\footnotesize{Effect of varying $U$ (fixed $\theta=0$).}}
        \label{fig:ub_analysis}
    \end{subfigure}

    \caption{Sensitivity analysis of the proposed algorithms. The red markers ($U=1$) correspond to the exclusive-dispatch baseline and serve as a common starting point for FA (blue) and BA (green) strategies.}
    \label{fig:sensitivity_analysis}
\end{figure}

\smallskip
\xhdr{Sensitivity analysis \& non-myopic notifications.} \revcolor{We next perform a sensitivity analysis by investigating how system parameters---the opportunity-cost threshold ($\theta$) and the notification upper bound ($U$)---impact our results.

\Cref{fig:sensitivity_analysis}~(a) varies the threshold $\theta$ with $U=3$ fixed. As $\theta$ increases, packing becomes more conservative (effectively the degree of parallelization reduces as fewer drivers are notified), which increases average match time and decreases the number of finalized matches for every algorithm under both protocols. The effect on the average score, in contrast, depends on how the algorithm reacts to the threshold. \textsc{Opt} re-optimizes the notification set: when $\theta$ increases, it replaces a high-scoring but unreliable driver with a more reliable one to preserve the match probability, so its score is flat under FA and declines under BA. \textsc{Approx} instead trims its rounded set, dropping lower-score drivers while always keeping its highest-scoring driver; as $\theta$ grows it effectively notifies only that driver, which raises its score at the cost of slower matching.

\Cref{fig:sensitivity_analysis}~(b) varies the notification cap $U$ with $\theta=0$ fixed. As $U$ increases, packing becomes less conservative, which decreases average match time and increases the number of finalized matches for every algorithm under both protocols. Most of the gain is realized by $U=3$; beyond it, the marginal improvements in match time and match count are small. The effect on average score, in contrast, depends on how the algorithm uses the additional notification capacity. \revcolor{\textsc{Opt}} re-optimizes the notification set: as $U$ increases, its score rises under BA and remains essentially flat under FA, since it avoids adding lower-scoring drivers whose faster responses could reduce match quality. \textsc{Approx} instead expands the score-ordered prefix of its rounded set. Starting from $U=1$, where it retains only the highest-scoring driver, larger values of $U$ admit progressively lower-scoring drivers. Under FA, these drivers can respond first and effectively ``hijack'' the ride from a closer or more suitable driver, causing the average score to decline steadily. The decline is milder under BA because the platform selects the best acceptor, although the additional backups create lower-scoring matches in cases that would otherwise remain unmatched. These results suggest that intermediate values of $U$ provide a favorable speed--quality balance, particularly under FA.}

\smallskip
\xhdr{k-accept Protocol.}
\revcolor{Finally, we analyze the impact of different contention resolution protocols on global system performance. In \Cref{sec:prelim}, we introduced the $k$-accept protocol as a generalized mechanism that interpolates between the speed of FA and the score maximization of BA. 

Figure \ref{fig:K3} presents the simulation results for the optimal algorithm under varying $k \in \{1, 2, 3\}$, with capacity fixed at $U=3$ and threshold $\theta=0$. Note that the boundary case $k=1$ corresponds to the standard FA policy, while $k=3$ corresponds to the BA policy.

\begin{figure}[H]
    \centering    \includegraphics[width=0.9\linewidth]{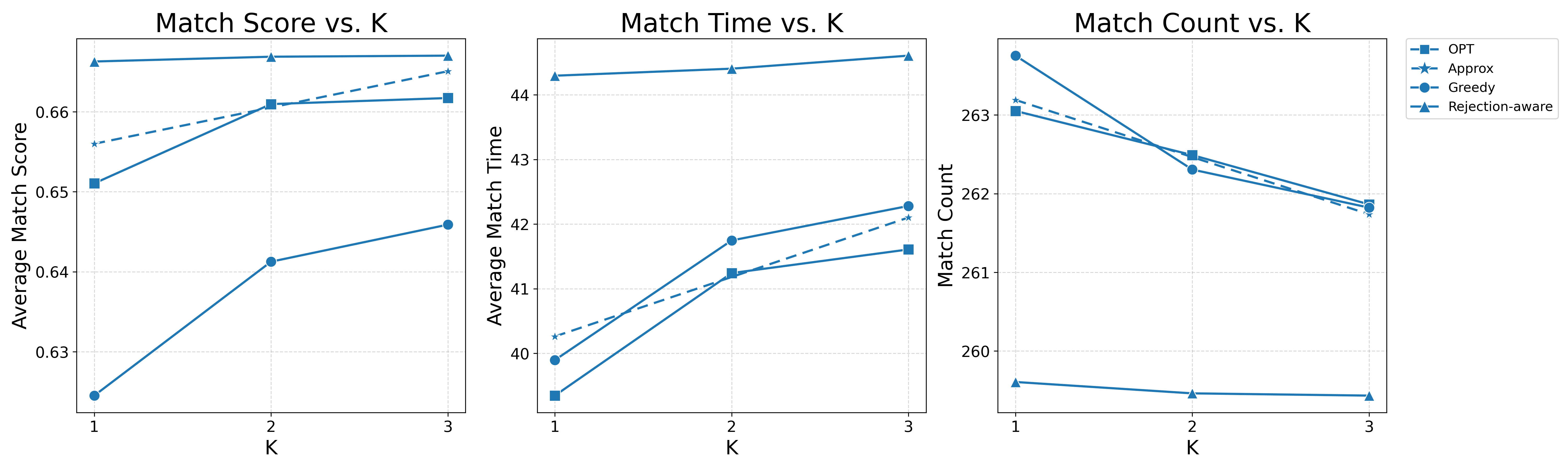}
    \caption{Performance metrics (average score, average match time, match count) for the optimal algorithm, i.e., NED OPT, under $k$-accept strategies with $\boldsymbol{U=3,\theta=0}$, for $\boldsymbol{k\in\{1,2,3\}}$.}
    \vspace{-3mm}
    \label{fig:K3}
\end{figure}

The results reveal a sharp trade-off between match quality and system latency. Shifting from $k=1$ (FA) to $k=2$ produces a significant jump in the average match score, validating the intuition that waiting for even a single additional response drastically improves the likelihood of finding a high-quality driver. However, this gain comes at a cost: both match time and total match count degrade noticeably. The increase in match time reflects the latency inherent in waiting for multiple drivers to respond, while the drop in total match count suggests that the stricter requirements of higher $k$ values may lead to more unfulfilled requests in a dynamic setting.

This confirms the theoretical flexibility of the $k$-Accept lever: lower values of $k$ prioritize throughput and speed (minimizing rider waiting time), while higher values prioritize the quality of individual matches at the expense of system fluidity/speed. As before, there again exists an inherent tradeoff between speed and quality.}

\subsection{Further Discussions}
\label{apx:simulation-discussion}
\begin{figure}[htb]
    \centering
    \includegraphics[width=0.95\linewidth]{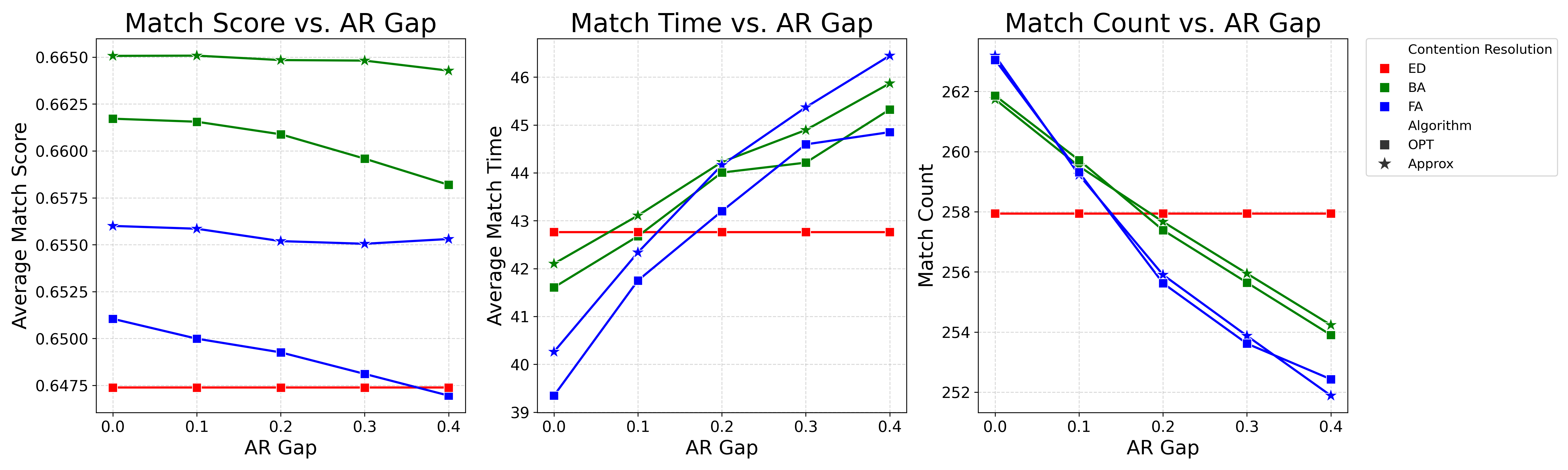}
    \caption{Performance impact of the driver AR-GAP on optimal NED policies. An AR-GAP of $\boldsymbol{x}$ scales the baseline ED acceptance probability $\boldsymbol{p}$ to $\boldsymbol{(1-x)p}$ under NED. While an AR-GAP of  $\boldsymbol{\approx 0.4}$ largely neutralizes NED's advantages, its gains are fully realized as the acceptance rates converge (around AR-GAP $\boldsymbol{\le 0.2}$).}
    \label{fig:AR}
\end{figure}

\xhdr{Drivers trust and the ``Acceptance Rate Gap (AR-GAP)''.}
\revcolor{An important consideration in} the transition to \revcolor{an} NED notification system is a behavioral feedback loop we term the \emph{Acceptance Rate Gap (AR-GAP)}. If an \FEedit{NED} system is implemented without safeguarding driver experience, it risks breaking trust in the platform's reliability. This erosion can lead to a decrease in drivers' acceptance probabilities, imposing significant long-term costs on the platform. Specifically, if drivers frequently encounter contention, i.e., situations in which they accept a request only to be rejected in favor of another driver, they may become reluctant to engage with future notifications. This frustration can precipitate a decline in actual acceptance rates, severely undermining the benefits of the new design. \FEedit{See \Cref{fig:AR} for an illustration.}

This adverse dynamic is particularly acute under the BA protocol, where a driver who successfully accepts a request is often explicitly rejected for a higher-utility candidate. In contrast, this issue is significantly mitigated under FA. In an FA system, the platform can retract outstanding notifications immediately upon the first acceptance, ensuring that contention events are rare and primarily driven by unavoidable network latency rather than protocol design.

We should also note that, as our study shows  empirically (\Cref{sec:simulation}) and we later confirm theoretically (\Cref{sec:MMM}), in the NED notification system the throughput is higher, resulting in increased drivers' earning opportunities. Therefore, while at the surface level a driver might feel that they are sometimes getting rejected, they will eventually be able to serve higher quality and in fact more number of riders---a win-win-win situation for the platform, drivers, and riders.

\smallskip
\xhdr{Implementation trade-offs: FA vs. BA.}
Beyond the behavioral risks associated with the AR-gap, the decision to implement FA versus BA requires a broader perspective. While our simulations quantify the theoretical trade-offs in long-run performance, the operational choice involves practical considerations that extend beyond pure algorithmic study but are important in practice:

\begin{itemize}
    \item \textit{Driver experience and fairness:} BA guarantees \emph{local} optimality, ensuring the rider is matched with the best driver among those who accepted. This quality-based logic is easier to justify to drivers than a speed-based criterion. However, BA imposes a mandatory wait time; losing a match after waiting for the window to close can be a frustrating user experience. FA eliminates this ``waiting in vain'' by providing instant feedback, allowing rejected drivers to immediately consider other options.
    
    \item \textit{Incentive structure and cancellations:} FA creates a ``race'' dynamic that incentivizes drivers to respond as quickly as possible. While this speeds up market clearing, it may encourage ``blind acceptance,'' where drivers accept a request immediately without scrutinizing the trip details. This behavior can lead to higher post-acceptance cancellation rates, as drivers may renege once they realize the trip does not meet their preferences, thereby disrupting service reliability. 
\end{itemize}

\smallskip
\xhdr{Towards a fully decentralized market design.}
The proposed NED framework represents a step away from purely centralized dispatching (where the platform dictates assignments) toward a semi-centralized market mechanism. However, a \emph{fully} decentralized solution would eliminate the packing layer entirely, allowing drivers to view and bid on any ride in the system, potentially regulated by dynamic pricing or congestion controls. Although our current approach optimizes the ``notification sets'' to maximize system welfare, future research might explore mechanisms where the platform acts solely as a clearinghouse, allowing the market equilibrium to emerge organically from driver-initiated filtering and selection.

\section{Marketplace Macro Model}
\label{sec:MMM}
In this section, we develop a stylized marketplace fluid model to isolate the long-run (i.e., equilibrium) implications of the FA and BA contention-resolution protocols. The model is intentionally minimal: it abstracts away spatial heterogeneity and other platform details to focus on two questions suggested by our large-scale simulations in \Cref{sec:simulation}: \emph{(i) Can we theoretically justify the inherent trade-off between speed (FA) and quality (BA)? (ii) Does the choice of FA versus BA systematically thicken or thin the market in steady state?}

Our theoretical analysis---as we elaborate throughout this section---highlights two countervailing forces, preventing either protocol from emerging as a dominant winner in terms of market thickness. Relative to BA, the FA protocol clears notifications more quickly, so drivers and riders spend less time in the notification stage; this reduces friction and tends to increase the mass of agents available in the idle pools. At the same time, because FA attains a higher throughput of successful matches, it sustains higher driver utilization and therefore depletes the stock of idle drivers faster than BA. These forces offset in steady state. Consistent with the simulation evidence in \Cref{sec:simulation}, equilibrium market thickness is comparable under both regimes; consequently, long-run performance is driven primarily by the intrinsic speed--versus--quality trade-off rather than by major differences in endogenous market size.

\smallskip
\xhdr{Model.} We consider a continuum model with mass dynamics on both sides of the market. At time $t$, there are $D_t$ idle drivers and $R_t$ waiting riders in the system. New drivers arrive at rate $\driverArrival$, joining the idle pool, and new riders arrive at rate $\riderTotalArrival$. At dispatch epochs (matching cycles), which occur according to a Poisson process, the platform selects a subset of at most $U$ idle drivers to notify for each waiting rider, following a given NED algorithm \textsc{ALG} for single-cycle. Each notified driver responds after an exponential time with rate $\mu$, independently accepting the request with probability $p$ (and rejecting it with probability $1-p$). Upon acceptance, the platform either finalizes the match immediately or continues waiting for additional responses (depending on contention resolution protocol). 

Riders leave the platform and seek alternative options at rate $\rideChurn$. This abandonment rate is independent of the notification state, since riders do not observe which drivers have been contacted until a match is finalized. Idle drivers also leave at rate $\idleChurn$, while notified drivers leave at a rate $\notifChurn$.

\smallskip
\xhdr{Large market fluid model.} To emphasize steady-state tradeoffs, we use a standard large-market scaling: all arrival, service, and churn rates are multiplied by a large factor $N$, and we study the resulting fluid limits rather than the full stochastic trajectories.~\footnote{Our fluid model is based on similar notions such as \emph{mean-field equilibrium} (e.g., see \cite{light2022mean}) or \emph{stationary equilibrium}~\citep{adlakha2015equilibria}. This approximation captures aggregate system behavior and allows us to characterize equilibrium driver availability and matching performance.} We focus on long-run averages over a horizon $T$ and summarize equilibrium market thickness by $(R_0, D_0)$, where $R_0$ denotes the long-run average number of riders awaiting notification, and $D_0$ denotes the long-run average number of idle drivers. Furthermore, let $q_i$ represent the steady-state fraction of riders who notify exactly $i$ drivers during a dispatch epoch. These quantities are strictly endogenous; they emerge from the interplay between the platform's packing algorithm and the contention resolution protocol. 

In Sections~\ref{sec:MMMFA} and \ref{sec:MMMBA}, we analyze these structural interactions under the FA and BA protocols, respectively, for a fixed dispatch profile $\{q_i\}$. Building on these derivations, \Cref{sec:equilibrium_method} details the methodology for computing the equilibrium market thickness, successful match rate, and expected match time. Finally, in \Cref{Sec:MMMSimulation}, we conduct numerical experiments to compare the protocols under this stylized framework, corroborating the extensive simulation results presented in \Cref{sec:simulation}.

\subsection{First Accept Fluid Model}\label{sec:MMMFA}
Under the FA rule, contention is resolved at the first acceptance, so the system dynamics simplify substantially. Each rider is either waiting to be notified to some drivers, or waiting for responses from a set of $i \in \{1,\dots,U\}$ notified drivers. Because the platform finalizes the match with the first accepting driver, riders never enter intermediate ``accepted'' states: when any of the $i$ drivers accepts, the rider leaves the system and all other notified drivers return to the idle pool.

\smallskip
\xhdr{Riders.}  Consider a rider with $i$ outstanding notifications. Responses arrive at total rate $\mu i$, and each response is an acceptance with probability $p$. Thus, the rider is matched at rate $\mu i p$. If a responding driver rejects, which occurs at rate $\mu i(1-p)$, the rider transitions to a state with $i-1$ outstanding notifications.  This creates a cascade of states $i \to i-1 \to \cdots \to 0$, where a rider either finds a match or, upon reaching state $0$, is returned to the pool of riders waiting to be notified again. From there, the platform may issue a new set of notifications, or the rider may eventually abandon the system.

\smallskip
\xhdr{Drivers.} The driver-side dynamics mirror these transitions. Each rider in state $i$ corresponds to exactly $i$ notified drivers in the dual driver state. A notified driver accepts at rate $\mu p$, in which case the match is finalized immediately and the driver leaves the system to serve the rider. A notified driver returns to the idle pool either by rejecting at rate $\mu(1-p)$, or because another notified driver accepts first, which occurs at rate $\mu(i-1)p$.

\begin{figure}[htb]
\centering
\resizebox{0.8\linewidth}{!}{%
\setlength{\tabcolsep}{2pt} 
\begin{tabular}{c|c} \textbf{\large Riders} & \textbf{\large Drivers}\\

\begin{tikzpicture}[
    node distance=1.5cm and 2cm,
    every node/.style={draw, minimum size=1cm, font=\tiny, align=center},
    arrow/.style={->, thick},
    round/.style={circle, draw, minimum size=1cm}
]

\node[round] (WaitingR) at (-4,0) {$R_0$};
\draw[arrow] (-5.2,0) -- (WaitingR)
  node[midway, above, draw=none, yshift = -8pt] {$\riderTotalArrival$};
\draw[arrow, color=red!80!black] (WaitingR)--(-4,1.2)
  node[midway, left, draw=none, xshift = 10pt] {$\rideChurn$};

\node[round] (R3) at (-1,3) {$R_3$};
\node[round] (R2) at (-1,0) {$R_2$};
\node[round] (R1) at (-1,-3) {$R_1$};

\draw[arrow] (R3) -- (R2) node[midway, draw=none,xshift = 27pt] {$3\left[\mu(1-p)+\notifChurn\right]$};
\draw[arrow] (R2) -- (R1) node[midway, draw=none,xshift = 27pt] {$2\left(\mu(1-p)+\notifChurn\right)$};
\draw[arrow, color=green!70!black] (R3) -- (0,3.5)
    node[midway, below, draw=none, yshift=12pt,xshift = 8pt] {$3\mu p$};
\draw[arrow, color=red!80!black] (R3) -- (0,2.5)
    node[midway, above, draw=none, yshift=-12pt,xshift = 8pt] {$\rideChurn$};
\draw[arrow, color=green!70!black] (R2) -- (0,0.5)
    node[midway, below, draw=none, yshift=12pt,xshift = 8pt]  {$2\mu p$};
\draw[arrow, color=red!80!black] (R2) -- (0,-0.5)
    node[midway, above, draw=none, yshift=-12pt,xshift = 8pt]  {$\rideChurn$};

\draw[arrow, color=green!70!black] (R1) -- (0,-2.5)
    node[midway, below, draw=none, yshift=12pt,xshift = 8pt]  {$\mu p$};
\draw[arrow, color=red!80!black] (R1) -- (0,-3.5)
    node[midway, above, draw=none, yshift=-12pt,xshift = 8pt]  {$\rideChurn$};

\draw[arrow] (WaitingR) -- (R1)
  node[midway, below, draw=none, yshift = 4pt] {$q_1$};
\draw[arrow] (WaitingR) -- (R2)
  node[midway, below, draw=none, yshift = 8pt] {$q_2$};
\draw[arrow] (WaitingR) -- (R3)
  node[midway, below, draw=none, yshift = 4pt] {$q_3$};

\draw[->, thick]
  (R1.west) arc[start angle=-90, end angle=-180, x radius=2.5cm, y radius=2.5cm] node[midway, below, draw=none, yshift = -2pt] {$\mu(1-p)+\notifChurn$};

\end{tikzpicture}
&
\begin{tikzpicture}[
    node distance=1.5cm and 2cm,
    every node/.style={draw, minimum size=1cm, font=\tiny, align=center},
    arrow/.style={->, thick},
    round/.style={circle, draw, minimum size=1cm}
]

\node[round] (WaitingD) at (4,0) {$D_0$};
\draw[arrow] (5.2,0) -- (WaitingD)
  node[midway, above, draw=none, yshift = -8pt] {$\driverTotalArrival$};
\draw[arrow, color=red!80!black] (WaitingD)--(5,-1)
  node[midway, left, draw=none, xshift = 8pt] {$\idleChurn$};

\node[round] (D3) at (1,3) {$D_3=3R_3$};
\node[round] (D2) at (1,0) {$D_2=2R_2$};
\node[round] (D1) at (1,-3) {$D_1=R_1$};

\draw[arrow] (D3) -- (D2) node[midway, left, draw=none, yshift=0pt,xshift = 0pt] {$2\mu (1-p)+2\notifChurn$};
\draw[arrow] (D2) -- (D1) node[midway, left, draw=none, yshift=0pt,xshift = 0pt] {$\mu (1-p)+\notifChurn$};

\draw[arrow, color=green!70!black] (D3) -- (0,3.5)
    node[midway, below, draw=none, yshift=12pt,xshift = -8pt] {$\mu p$};
\draw[arrow, color=red!80!black] (D3) -- (0,2.5)
    node[midway, above, draw=none, yshift=-12pt,xshift = -8pt] {$\notifChurn$};
\draw[arrow, color=green!70!black] (D2) -- (0,0.5)
    node[midway, below, draw=none, yshift=12pt,xshift = -8pt] {$\mu p$};
\draw[arrow, color=red!80!black] (D2) -- (0,-0.5)
    node[midway, above, draw=none, yshift=-12pt,xshift = -8pt] {$\notifChurn$};
\draw[arrow, color=green!70!black] (D1) -- (0,-2.5)
    node[midway, below, draw=none, yshift=12pt,xshift = -8pt] {$\mu p$};
\draw[arrow, color=red!80!black] (D1) -- (0,-3.5)
    node[midway, above, draw=none, yshift=-12pt,xshift = -8pt] {$\notifChurn$};

\draw[arrow] (WaitingD) -- (D1)
  node[midway, below, draw=none, yshift = 4pt] {$q_1\frac{R_0}{D_0}$};
\draw[arrow] (WaitingD) -- (D2)
  node[midway, above, draw=none, yshift = -8pt] {$2q_2\frac{R_0}{D_0}$};
\draw[arrow] (WaitingD) -- (D3)
  node[midway, below, draw=none,xshift= -2pt, yshift = 4pt] {$3q_3\frac{R_0}{D_0}$};

\draw[->, thick]
  (D1.east) arc[start angle=-90, end angle=0, x radius=2.5cm, y radius=2.5cm] 
  node[midway, below, draw=none, yshift = -2pt] {$\mu(1-p)+\rideChurn$};

\draw[->, thick]
  ($(D2.south east)!0.8!(D2.east) + (0.5pt,0) $)
    arc[start angle=-135, end angle=-60, x radius=1.6cm, y radius=1.5cm]
    node[midway, below, draw=none, xshift=-12pt, yshift=10pt] {$\mu p +\mu(1-p)+\rideChurn$};

\draw[->, thick]
  (D3.east) arc[start angle=90, end angle=0, x radius=2.5cm, y radius=2.5cm] node[midway, below, draw=none, xshift = 24pt, yshift = 20pt] {$2\mu p +\mu(1-p)+\rideChurn$};

\end{tikzpicture}

\end{tabular}%
}

\caption{State transition diagram for the FA protocol with $\boldsymbol{U=3}$.}
\label{fig:FAMMM}
\end{figure}
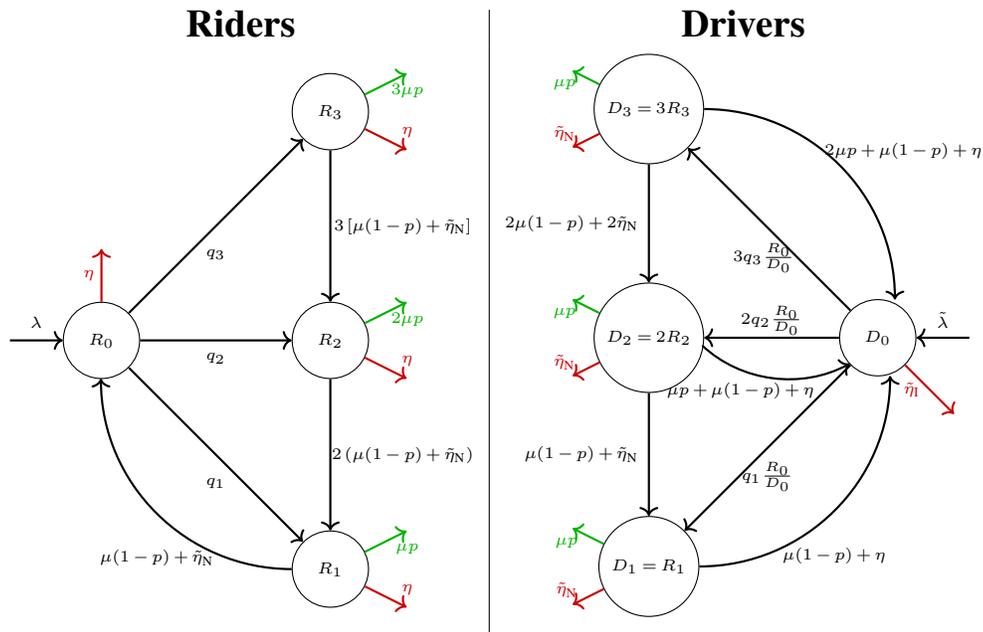

Finally, without loss of generality, we normalize the notification rate to one, so that a rider spends one unit of time in the idle state before being (re)notified to drivers.
\Cref{fig:FAMMM} illustrates these flows for $U=3$. The resulting fluid dynamics describe how mass moves across rider states and driver states under FA, and will allow us to characterize both steady-state availability and matching throughput

Let $R_i$ denote the mass of riders currently waiting on an active notification set of size $i$, and $D_i$ the corresponding mass of drivers processing these requests. By construction, since every rider in state $R_i$ has contacted exactly $i$ drivers, the aggregate mass of drivers in the dual state satisfies the coupling condition $D_i = i R_i$, as in \Cref{fig:FAMMM}. This dynamic is governed by the following system of \emph{flow equations} (or equivalently, steady-state equations for the corresponding CTMC):
\begin{align}
\tag{\textsc{FA-Equilibrium}}
    R_{\ell} \left( \rideChurn + \ell(\mu+\notifChurn) \right) &= R_0 q_{\ell} + (\ell+1) R_{\ell+1}\left(\mu (1-p) + \notifChurn\right) && 1\leq \ell \leq U \\
    R_0\left(\rideChurn + \sum_{\ell=1}^U q_{\ell}\right) &= \riderTotalArrival + R_1\left(\mu(1-p)+\notifChurn\right)\nonumber\\
    D_0\idleChurn + R_0\sum_{\ell=1}^U \ell q_{\ell} &= \driverTotalArrival     + \sum_{\ell = 1}^U D_{\ell} \left(\rideChurn + \mu(1-p) +(\ell-1)\mu p \right)~,\nonumber
\end{align}
where, as a convention, we define $R_{U+1}$ to be zero. Solving these steady-state equilibrium equations characterizes the market sizes $(R_0, D_0)$ as a function of the dispatch probabilities $\{q_1,\dots,q_U\}$.

\begin{proposition}[FA Equilibrium]
\label{prop:FA}
    The long-run equilibrium market sizes under FA is given by:
    \begin{align*}
        R_0 = \frac{\lambda}{\rideChurn + \sum_{i=1}^U q_i\left(1-\prod_{j=1}^i \frac{j \left(\mu(1-p)+\notifChurn\right)}{\rideChurn + j (\mu+\notifChurn)}\right)},\quad
        \frac{\driverTotalArrival- D_0\idleChurn}{\mu p+ \notifChurn} =  \frac{\frac{\riderTotalArrival}{\mu(1-p)+\notifChurn}\sum_{\ell = 1}^U \left( \sum_{i=\ell}^{U}q_{i} \prod_{j=\ell}^i \frac{j \left(\mu(1-p)+\notifChurn\right)}{\rideChurn + j (\mu+\notifChurn)}\right)}{\rideChurn + \sum_{i=1}^U q_i\left(1-\prod_{j=1}^i \frac{j \left(\mu(1-p)+\notifChurn\right)}{\rideChurn + j (\mu+\notifChurn)}\right)}.
    \end{align*}
\end{proposition}
While the specific algebraic form of these equations is secondary, the crucial insight is structural: there exists a continuous mapping from the notification size distribution $\{q_i\}$ to the steady-state market thickness $(R_0, D_0)$. Conversely, the platform's dispatch logic (via the packing algorithm) induces a continuous mapping from the current market state $(R_0, D_0)$ back to the realized notification frequencies $\{q_i\}$. Consequently, the composition of these functions defines a continuous map from the probability simplex to itself. By Brouwer's Fixed Point Theorem, there exists a fixed point representing the equilibrium of the fluid system.

\begin{example}  When $\rideChurn = 0$, the expressions simplify significantly. Let $\alpha \coloneqq \frac{\mu(1-p)+\notifChurn}{\mu+\notifChurn}$. Then:
\begin{align*}
    \frac{\driverTotalArrival- D_0\idleChurn}{\mu p+\notifChurn}
    = \frac{\frac{\riderTotalArrival}{\mu(1-p)+\notifChurn}\sum_{\ell = 1}^U \left( \sum_{i=\ell}^{U}q_{i} \prod_{j=\ell}^i \alpha\right)}{\sum_{i=1}^U q_i\left(1-\prod_{j=1}^i \alpha\right)}
    = \frac{\frac{\riderTotalArrival}{\mu+\notifChurn}\sum_{i=1}^{U}q_{i}\left( \sum_{\ell = 1}^{i} \alpha^{i-\ell} \right)}{\sum_{i=1}^U q_i\left(1-\alpha^i\right)}
    = \frac{\riderTotalArrival}{\mu p} .
\end{align*}
Observe that the expression derived above simplifies to the constant $\frac{\riderTotalArrival}{\mu p}$, rendering it invariant with respect to the notification frequencies $\{q_i\}$. Consequently, in this idealized regime, the specific parameterization of the dispatch heuristic—whether \FEedit{ED, NED}, or varying degrees of myopia—has no impact on the equilibrium $D_0$. However, in practical settings where rider patience is finite ($\rideChurn > 0$), this invariance breaks down. The presence of abandonment explicitly couples the market thickness to the platform's dispatch policy.
\end{example}

\subsection{Best Accept Fluid Model}\label{sec:MMMBA}

Under the BA rule, contention is resolved only after the platform collects responses from all notified drivers and selects the best available acceptance. As a result, the system dynamics are richer than under FA. Each rider is either waiting to be notified to drivers, waiting for responses from a set of $i \in \{1,\dots,U\}$ notified drivers, or is in an intermediate accepted state in which one or more drivers have accepted but the platform has not yet finalized the match. Unlike FA, a rider does not immediately leave the system upon the first acceptance, since the platform may wait to compare multiple acceptances before assigning the ride.

\smallskip
\xhdr{Riders.} Consider a rider who has been notified to $i$ drivers. Responses from these drivers arrive at total rate $\mu i$. A rejection reduces the number of outstanding notifications, transitioning the rider from state $i$ to ${i-1}$. An acceptance, however, moves the rider into an accepted state rather than immediately finalizing the match. While in an accepted state, the rider may continue to receive additional responses from the remaining notified drivers, potentially accumulating multiple acceptances. The platform then finalizes the match according to the BA rule, selecting the best accepting driver once all the drivers responded. As riders do not observe this process, they depart at the rate $\rideChurn$ regardless of state.

If all notified drivers reject, the rider eventually reaches state $0$ and returns to the pool of riders waiting to be notified again. Under BA, once the $i$-th best notified driver accepts, notifications to all inferior drivers can be withdrawn since they will never be selected. At that moment, the rider enters an accepted state $A_i$ (we abuse the notation and use $A_i$ to denote both the accepted state $i$ for the rider and also the mass of riders at this state), in which exactly $i-1$ higher-ranked notified drivers remain outstanding. While in $A_i$, if an additional driver $j<i$ accepts, the rider transitions to $A_j$, reflecting an improved best available acceptance. If one of the remaining higher-ranked drivers rejects, the rider transitions to $A_{i-1}$.

Thus, accepted states evolve dynamically as responses arrive: acceptances induce transitions across accepted states corresponding to better ranked drivers, while rejections reduce the number of remaining candidates. This structure captures the key feature of BA, namely that the platform retains the best acceptance observed so far while continuing to wait for potentially better ones.

\xhdr{Drivers.} The driver-side dynamics reflect this delayed resolution. A driver notified for a given ride responds at rate $\mu$. Upon acceptance, the driver does not immediately leave the system, but instead enters an intermediate accepted state where they remain pending until the platform finalizes the assignment. 
An accepting driver may be released back to the idle pool if another accepting driver is chosen, or if the rider exits the system. Rejections immediately return the driver to the idle pool.

These dynamics are modeled using $2U$ states. Let $D_0$ denote the mass of idle drivers who are not currently processing any notifications. For $i \in \{1, \dots, U\}$, let $D_i$ represent the set of drivers who have been notified for a request and hold the $i$-th rank among all candidates notified for that specific ride. This formulation highlights a fundamental structural difference from the FA protocol. In the FA setting, a driver's state is determined by the \emph{quantity} of competitors (the total number of notified drivers). In contrast, under the BA protocol, the state is determined by the driver's \emph{relative rank} (ordinality) within the notified set, as this ranking dictates the priority order for match finalization.

In addition, for each $i \in \{2,\dots,U\}$, we introduce an accepted state $\tilde{A}_i$, which consists of drivers who have accepted a ride and currently hold the $i$-th best score for that rider. Drivers in $\tilde{A}_i$ remain pending until the platform either receives a better acceptance, in which case they are released back to $D_0$, or a rejection from a higher-ranked driver, in which case they transition to state $\tilde{A}_{i-1}$. 

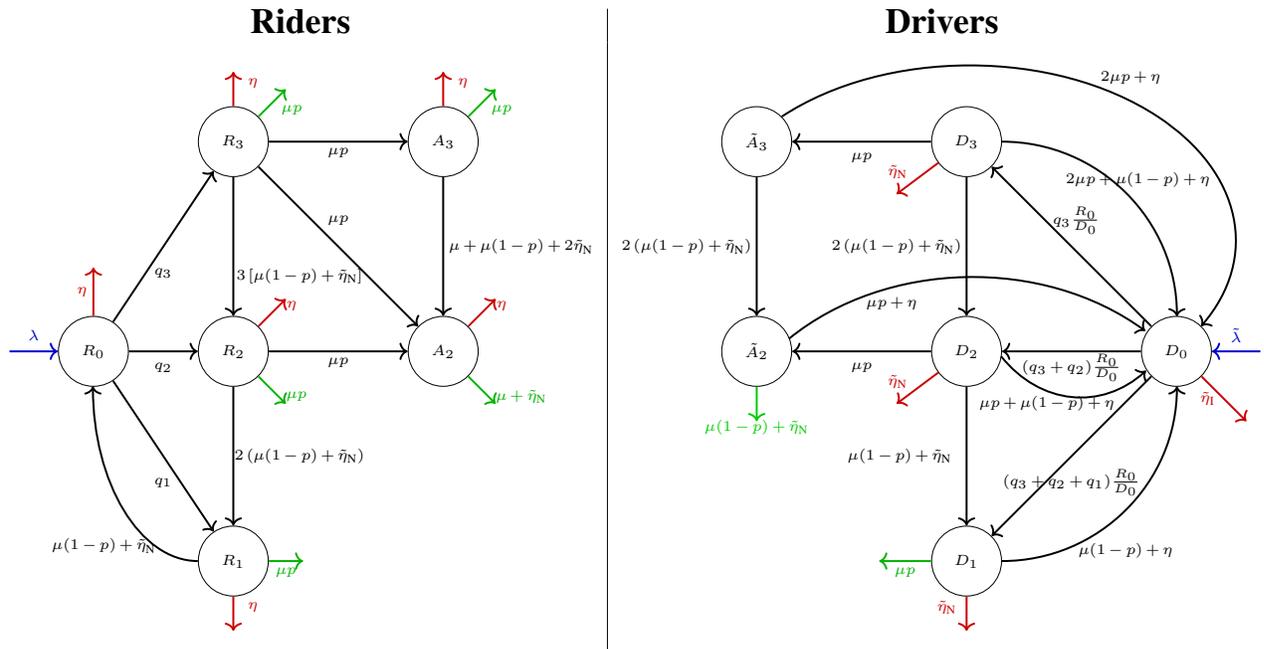
\begin{figure}[htb]
\centering
\setlength{\tabcolsep}{2pt} 
\scalebox{0.93}{
\begin{tabular}{c|c} \textbf{\large Riders} & \textbf{\large Drivers}\\

\begin{tikzpicture}[
    node distance=1.5cm and 2cm,
    every node/.style={draw, minimum size=1cm, font=\tiny, align=center},
    arrow/.style={->, thick},
    round/.style={circle, draw, minimum size=1cm}
]

\node[round] (WaitingR) at (-3,0) {$R_0$};
\draw[arrow, color=blue!80!black] (-4.2,0) -- (WaitingR)
  node[midway, above, draw=none, yshift = -8pt] {$\riderTotalArrival$};
\draw[arrow, color=red!80!black] (WaitingR)--(-3,1.2)
  node[midway, left, draw=none, xshift = 10pt] {$\rideChurn$};

\node[round] (R3) at (-1,3) {$R_3$};
\node[round] (R2) at (-1,0) {$R_2$};
\node[round] (R1) at (-1,-3) {$R_1$};

\node[round] (A3) at (2,3) {$A_3$};
\node[round] (A2) at (2,0) {$A_2$};

\draw[arrow] (R3) -- (R2) node[midway, draw=none,yshift = -12pt, xshift = 27pt] {$3\left[\mu(1-p)+\notifChurn\right]$};
\draw[arrow] (R2) -- (R1) node[midway, draw=none,xshift = 27pt] {$2\left(\mu(1-p)+\notifChurn\right)$};
\draw[arrow, color=green!70!black] (R3) -- (-0.25,3.75)
    node[midway, below, draw=none, yshift=12pt,xshift = 8pt] {$\mu p$};
\draw[arrow, color=red!80!black] (R3) -- (-1,4)
    node[midway, above, draw=none, yshift=-12pt,xshift = 8pt] {$\rideChurn$};
\draw[arrow, color=green!70!black] (R2) -- (-0.25,-0.75)
    node[midway, below, draw=none, yshift=12pt,xshift = 10pt]  {$\mu p$};
\draw[arrow, color=red!80!black] (R2) -- (-0.25,0.75)
    node[midway, above, draw=none, yshift=-12pt,xshift = 8pt]  {$\rideChurn$};

\draw[arrow, color=green!70!black] (R1) -- (0,-3)
    node[midway, below, draw=none, yshift=10pt]  {$\mu p$};
\draw[arrow, color=red!80!black] (R1) -- (-1,-4)
    node[midway, above, draw=none, yshift=-12pt,xshift = 8pt]  {$\rideChurn$};

\draw[arrow] (WaitingR) -- (R1)
  node[midway, below, draw=none, yshift = 4pt] {$q_1$};
\draw[arrow] (WaitingR) -- (R2)
  node[midway, below, draw=none, yshift = 8pt] {$q_2$};
\draw[arrow] (WaitingR) -- (R3)
  node[midway, below, draw=none, yshift = 4pt] {$q_3$};

\draw[arrow] (R2) -- (A2)
  node[midway, below, draw=none, yshift = 10pt] {$\mu p$};
\draw[arrow] (R3) -- (A3)
  node[midway, below, draw=none, yshift = 10pt] {$\mu p$};
\draw[arrow] (R3) -- (A2)
  node[midway, above, draw=none, yshift = -4pt] {$\mu p$};
  
\draw[arrow] (A3) -- (A2)
  node[midway, left, draw=none, xshift = 65pt] {$\mu + \mu(1-p)+2\notifChurn$};
  
\draw[arrow, color=green!70!black] (A3) -- (2.75,3.75)
    node[midway, below, draw=none, yshift=12pt,xshift = 8pt] {$\mu p$};
\draw[arrow, color=red!80!black] (A3) -- (2,4)
    node[midway, above, draw=none, yshift=-12pt,xshift = 8pt] {$\rideChurn$};
\draw[arrow, color=green!70!black] (A2) -- (2.75,-0.75)
    node[midway, below, draw=none, yshift=12pt,xshift = 16pt]  {$\mu + \notifChurn$};
\draw[arrow, color=red!80!black] (A2) -- (2.75,0.75)
    node[midway, above, draw=none, yshift=-12pt,xshift = 8pt]  {$\rideChurn$};

\draw[->, thick]
  (R1.west) arc[start angle=-90, end angle=-180, x radius=1.5cm, y radius=2.5cm] node[midway, below, draw=none, xshift = -8pt, yshift = 0pt] {$\mu(1-p)+\notifChurn$};

\end{tikzpicture}
&
\begin{tikzpicture}[
    node distance=1.5cm and 2cm,
    every node/.style={draw, minimum size=1cm, font=\tiny, align=center},
    arrow/.style={->, thick},
    round/.style={circle, draw, minimum size=1cm}
]

\node[round] (WaitingD) at (4,0) {$D_0$};
\draw[arrow, color=blue!80!black] (5.2,0) -- (WaitingD)
  node[midway, above, draw=none, yshift = -8pt] {$\driverTotalArrival$};
\draw[arrow, color=red!80!black] (WaitingD)--(5,-1)
  node[midway, left, draw=none, xshift = 8pt] {$\idleChurn$};

\node[round] (D3) at (1,3) {$D_3$};
\node[round] (D2) at (1,0) {$D_2$};
\node[round] (D1) at (1,-3) {$D_1$};

\node[round] (A3) at (-2,3) {$\tilde{A}_3$};
\node[round] (A2) at (-2,0) {$\tilde{A}_2$};

\draw[->, thick]
  (A3.north east) arc[start angle=135, end angle=-29.5, x radius=3.8cm, y radius=2.5cm] node[midway, above, draw=none, xshift = 0pt, yshift = -5pt] {$2\mu p +\rideChurn$};

\draw[->, thick]
  ($(A2.north east)+(3pt,-5pt)$) arc[start angle=135, end angle=48, x radius=3.7cm, y radius=3cm] node[midway, below, draw=none, xshift = -30pt, yshift = 4pt] {$\mu p+\rideChurn$};
\draw[arrow] (D3) -- (D2)
    node[midway, left, draw=none,xshift = 2pt] {$2\left(\mu (1-p)+\notifChurn\right)$};
\draw[arrow] (D2) -- (D1)
    node[midway, left, draw=none,xshift = -2pt] {$\mu (1-p)+\notifChurn$};
\draw[arrow] (D3) -- (A3)
    node[midway, below, draw=none,yshift = 8pt] {$\mu p$};
\draw[arrow] (D2) -- (A2)
    node[midway, below, draw=none,yshift = 8pt] {$\mu p$};
\draw[arrow] (A3) -- (A2)
    node[midway, left, draw=none,xshift = 2pt] {$2\left(\mu (1-p)+\notifChurn\right)$};
\draw[arrow, color=green!80!black] (A2) -- (-2,-1)
    node[midway, above, draw=none, yshift=-24pt] {$\mu(1-p)+\notifChurn$};

\draw[arrow, color=red!80!black] (D3) -- (0,2.25)
    node[midway, above, draw=none, yshift=-12pt,xshift = -8pt] {$\notifChurn$};
\draw[arrow, color=red!80!black] (D2) -- (0,-0.75)
    node[midway, above, draw=none, yshift=-12pt,xshift = -8pt] {$\notifChurn$};
\draw[arrow, color=green!70!black] (D1) -- (-0.25,-3)
    node[midway, below, draw=none, yshift=10pt,xshift = 0pt] {$\mu p$};
\draw[arrow, color=red!80!black] (D1) -- (1,-4)
    node[midway, above, draw=none, yshift=-12pt,xshift = -8pt] {$\notifChurn$};

\draw[arrow] (WaitingD) -- (D1)
  node[midway, below, draw=none, yshift = 4pt] {$(q_3+q_2+q_1)\frac{R_0}{D_0}$};
\draw[arrow] (WaitingD) -- (D2)
  node[midway, below, draw=none, yshift = 8pt] {$(q_3+q_2)\frac{R_0}{D_0}$};
\draw[arrow] (WaitingD) -- (D3)
  node[midway, above, draw=none,xshift= 2pt, yshift = -4pt] {$q_3\frac{R_0}{D_0}$};

\draw[->, thick]
  (D1.east) arc[start angle=-90, end angle=0, x radius=2.5cm, y radius=2.5cm] 
  node[midway, below, draw=none, yshift = -2pt] {$\mu(1-p)+\rideChurn$};

\draw[->, thick]
  ($(D2.south east)!0.8!(D2.east) + (0.5pt,0) $)
    arc[start angle=-135, end angle=-54, x radius=1.6cm, y radius=2cm]
    node[midway, below, draw=none, xshift=-10pt, yshift=12pt] {$\mu p +\mu(1-p)+\rideChurn$};

\draw[->, thick]
  (D3.east) arc[start angle=90, end angle=0, x radius=2.5cm, y radius=2.5cm] node[midway, below, draw=none, xshift = 5pt, yshift = 20pt] {$2\mu p +\mu(1-p)+\rideChurn$};

\end{tikzpicture}
\end{tabular}
}
\caption{State transition diagram for the BA protocol with $\boldsymbol{U=3}$.}
\label{fig:BAMMM}
\end{figure}

This state representation captures the essential feature of BA on the driver side: acceptances are retained and ranked, but only the best available acceptance ultimately leads to a match, while inferior acceptances may be revoked. As in the FA model, we normalize the notification rate to one, so that riders spend one unit of time in the idle state $R_0$ before being notified to drivers (if notified). \Cref{fig:BAMMM} illustrates the resulting state transitions for $U=3$. These dynamics form the basis for the fluid model under BA, which we use to analyze steady state behavior and matching performance.

Similar to FA dynamics, the driver states closely mirror the rider states. In particular, for each rider in state $R_i$, there is exactly one driver in each driver state $D_j$ for $j \le i$. Also for each, rider in state $A_i$ there are one driver in $\tilde{A_i}$ and one in each $D_j$ for $j < i$. More formally (by an abuse of notation to use the name of each state to also denote the mass of riders/drivers at that state),
\begin{align*}
    \tilde{A}_i = A_i, \qquad
    D_i = \sum_{j=i}^U R_j + \sum_{j=i+1}^U \tilde{A}_j
\end{align*}
Furthermore, we have the following system of flow equations characterizing an equilibrium: 

\begin{align}
 \raisetag{7.5ex}\tag{\textsc{BA-Equilibrium}}\label{eq:BA}
& R_{\ell}\left(\rideChurn+\ell(\mu+\notifChurn)\right)&&= R_0  q_{\ell} + (\ell+1) R_{\ell+1}\left(\mu (1-p) + \notifChurn\right)  &&& 1\leq \ell \leq U \\
& R_0\left(\rideChurn + \sum_{\ell=1}^U q_{\ell}\right)&&= \riderTotalArrival + R_1\left(\mu(1-p)+\notifChurn\right)\nonumber\\
& A_{\ell}\left(\rideChurn+(\ell-1)(\mu+\notifChurn)\right)&&= \mu p R_{\ell} + \ell A_{\ell+1}\left(\mu (1-p)+ \notifChurn\right) + \mu p\sum_{i=\ell+1}^U (A_i+R_i) &&&  2\leq \ell \leq U \nonumber\\
& D_0\idleChurn + R_0\sum_{\ell=1}^U \ell q_{\ell}&&= \driverTotalArrival     + \sum_{\ell = 1}^U D_{\ell} \left(\rideChurn + \mu(1-p) + (\ell-1)\mu p \right)+ \sum_{\ell = 2}^U \tilde{A}_{\ell} \left(\rideChurn + (\ell-1)  \mu p\right)~,\nonumber
\end{align}

\noindent where, for notational convenience, we define $R_{U+1}=A_{U+1} = 0$. 

\begin{proposition}[BA Equilibrium]
\label{prop:BA}
    The long-run equilibrium under BA is given by:
    \begin{align*}
        R_0 &= \frac{\lambda}{\rideChurn + \sum_{i=1}^U q_i\left(1-\prod_{j=1}^i \frac{j \left(\mu(1-p)+\notifChurn\right)}{\rideChurn + j (\mu+\notifChurn)}\right)} \\ 
        A_{\ell} &= R_0\left(\frac{\sum_{i=\ell-1}^Uq_i\left(1-\prod_{j=\ell}^i \frac{j \left(\mu(1-p)+\notifChurn\right)}{\rideChurn + j (\mu+\notifChurn)}\right)}{\rideChurn + (\ell-1)(\notifChurn + \mu)}- \sum_{j=\ell}^U\left[\frac{\frac{\rideChurn+ (\ell-1)\mu p}{\rideChurn + (\ell-1)(\notifChurn + \mu)}\left(\prod_{i=\ell}^{j}\frac{i(\mu(1-p)+\notifChurn)}{\rideChurn + i(\notifChurn + \mu)}\right)\sum_{i=j}^Uq_i}{j(\mu(1-p)+\notifChurn)}\right]\right)\\
    D_{\ell} &= R_0\sum_{j=\ell}^U \left[ \frac{ \left(\prod_{i=\ell}^{j} \frac{i(\mu(1-p)+\notifChurn)}{\rideChurn + i(\notifChurn + \mu)}\right)\sum_{i=j}^Uq_i}{j(\mu(1-p)+\notifChurn)}\right]\\
    D_0&= \frac{\driverTotalArrival + \sum_{\ell = 1}^U D_{\ell} \left(\rideChurn + \mu(1-p) + (\ell-1)\mu p \right)+ \sum_{\ell = 2}^U \tilde{A}_{\ell} \left(\rideChurn + (\ell-1)  \mu p\right) - R_0\sum_{\ell=1}^U \ell q_{\ell}}{\idleChurn}.
    \end{align*}
\end{proposition}
It may appear puzzling that the \emph{formulas} for total mass of riders in the \(R_i\) states, including \(R_0\), coincides with that under FA. The reason is that, under both FA and BA, once a ride receives an acceptance it will eventually exit the platform and never return. Under FA, the ride is always matched to the first accepting driver. Under BA, however, the ride may be assigned to a better ranked driver or may renege while waiting, thereby leaving the platform without a match. Although the resulting balance equations are identical on the rider side, the size of the idle driver pool \(D_0\) is different. This, in turn, changes the set of feasible matching options for each ride, altering the notification probabilities \(q_i\) and thereby changing the equilibrium values of the \(R_i\) states, and in particular \(R_0\).

As in the FA case, provided the dispatch algorithm maps 
the current market size $(R_0,D_0)$ continuously to the notification profile $\mathbf{q}$, the system dynamics satisfy the conditions of Brouwer's Fixed Point Theorem. This guarantees the existence of at least one equilibrium in this fluid system.

\subsection{Equilibrium Analysis: The Iterative Fixed-Point Method}
\label{sec:equilibrium_method}
\begin{figure}[htbp]
    \centering
    \begin{tikzpicture}[
        scale=0.85,
        transform shape,
        node distance = 2cm and 1.5cm,
        auto,
        block/.style = {rectangle, draw, text width=14em, text centered, rounded corners, minimum height=4em},
        cloud/.style = {draw, ellipse, text width=7em, text centered, minimum height=6em},
        line/.style = {draw, -{Latex[length=3mm, width=2mm]}, thick}
    ]

    \node [block] (init_q) {Current Assumed Probabilities\\ $\mathbf{q} = (q_0, q_1, q_2, q_3)$};

    \node [block, below=of init_q, yshift = 0.4cm] (theory) {Theoretical State Calculation\\ (Fluid Model Flow Equations)};

    \node [block, right=of theory,xshift= +1.0cm] (simulation) {Simulation 
    Environment\\ \footnotesize - Sample 1000 instances based on $R_0, D_0$\\ - Run Algorithm};

    \node [block, right=of init_q ,xshift= +1.0cm] (measured_q) {Observed Frequencies\\ $\mathbf{q}' = (q'_0, q'_1, q'_2, q'_3)$};

    \node [cloud, left=of theory, node distance=1.5cm,yshift= +0cm, font=\footnotesize] (metrics) {Theoretical Metrics\\ (Match Time, Prob.)};

    \path [line] (init_q) -- (theory);

    \path [line] (theory) -- node[below] {$(R_0, D_0)$} (simulation);

    \path [line, dashed] (theory) -- (metrics);

    \path [line] (simulation) -- (measured_q);

    \path [line] (measured_q) -- node[above, font=\small] {Repeat until} node[below] { $\mathbf{q} \approx \mathbf{q}'$} (init_q);

    \end{tikzpicture}
    \caption{Schematic representation of the iterative process used to find the equilibrium in the stylized model.}
    \label{fig:stylized_model_cycle}\vspace{-2mm}
\end{figure}
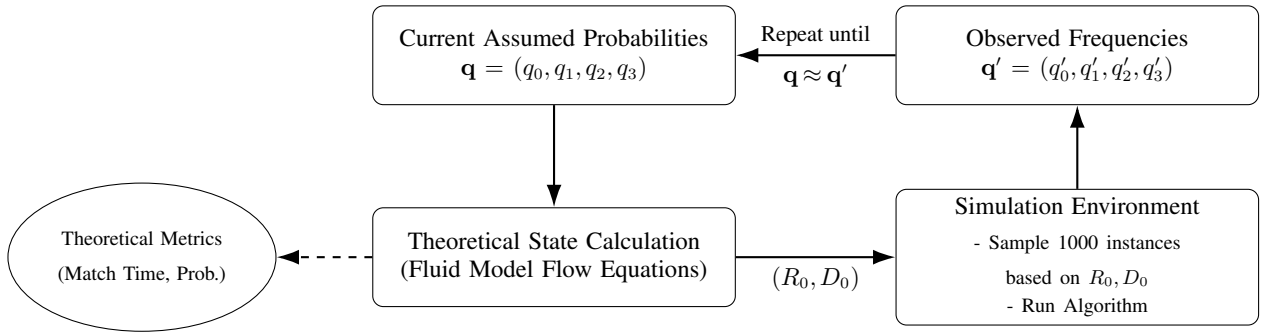

The fluid derivations in \Cref{sec:MMMFA,sec:MMMBA} treat the notification profile $\mathbf{q} = \{q_0, q_1, \dots, q_U\}$ as an input, where $q_i$ represents the probability that a randomly arriving rider is notified by exactly $i$ drivers. However, these probabilities are not exogenous constants; they are endogenous outcomes determined by the interplay between market density and the specific dispatch algorithm employed. This creates a circular dependency:
\begin{enumerate}
    \item The steady-state market sizes ($R_0, D_0$) depend on the match probabilities $\mathbf{q}$ (via the flow balance equations derived in \cref{sec:MMMFA} and \cref{sec:MMMBA} ).
    \item The actual performance of the dispatch algorithm and the resulting match probabilities depend on the density of available agents in the steady-state of the market ($R_0, D_0$).
\end{enumerate}
To resolve this interdependency, we compute a \emph{fixed point} $\mathbf{q}^*$ that corresponds to a consistent operating state in which algorithmic outcomes align with the steady-state assumptions. In particular, we employ an iterative fixed-point approach illustrated in Figure~\ref{fig:stylized_model_cycle}. The process is iterative, where each iteration is as follows:

\smallskip
\textit{1. Initialization \& analytical mapping.}
Starting from an initial guess $\mathbf{q}^{(t)}$ at the beginning of iteration $t$, we use the fluid flow equations in our analytical framework (\Cref{prop:FA} and \Cref{prop:BA}) to compute the implied steady-state market sizes, denoted $R_0(\mathbf{q}^{(t)})$ and $D_0(\mathbf{q}^{(t)})$. 

\smallskip
\textit{2. Simulation \& empirical measurement.}
To validate the belief $\mathbf{q}^{(t)}$, we simulate a marketplace snapshot with the computed densities $(R_0, D_0)$.
We generate 1,000 independent problem instances using the data generation process described in \Cref{Sec:MMMSimulation}, and then run the dispatch algorithm using the discrete-event simulator in \Cref{sec:simulation} (now with synthetic data) to record the outcomes. 

\smallskip
\textit{3. Update \& convergence.}
By aggregating the decisions made by the algorithm across all simulation instances, we obtain an empirical frequency vector $\mathbf{q}'$. If $\mathbf{q}^{(t)}$ matches the empirical frequencies (i.e., $\mathbf{q}^{(t)} \approx \mathbf{q}'$), we declare convergence. Otherwise, we update our belief $\mathbf{q}^{(t+1)}$ and go to the next iteration.

In the following Theorem we utilize the closed form solutions and prove that there exists an equilibrium point in this procedure.
\begin{theorem}\label{thm:existance}
    Let $\Delta^U = \{\mathbf{q} \in \mathbb{R}^{U+1} \mid \sum_{i=0}^U q_i = 1, q_i \geq 0\}$ denote the probability simplex of all valid notification profiles. Under the iterative procedure described in \Cref{fig:stylized_model_cycle}, there exists at least one fixed point $\mathbf{q}^* \in \Delta^U$ such that the expected empirical frequency vector $\mathbf{q}'$ generated by the simulation environment exactly matches the assumed theoretical profile $\mathbf{q}^*$.
\end{theorem}
\revcolor{
\begin{proofof}{Proof of Theorem~\ref{thm:existance}}
    \revcolor{Let} $\mathbf{e}_0=(1,0,\ldots,0)\in\Delta^U$ denote the notification profile under which no rider is notified to a driver. Let ${\Phi}: \Delta^U \to \Delta^U$ denote the composite mapping from the assumed {notification} profile $\mathbf{q}$ to the expected empirical profile $\mathbf{q}'$, such that ${\Phi}(\mathbf{q}) = \mathbf{q}'$. We can decompose this mapping into two distinct steps: the analytical mapping ${\Gamma}: \mathbf{q} \mapsto (R_0, D_0)$ and the simulation mapping ${\Psi}: (R_0, D_0) \mapsto \mathbf{q}'$. 

    The domain $\Delta^U$ is a probability simplex, which is by definition a compact and convex subset of $\mathbb{R}^{U+1}$. In previous sections, we established that the analytical mapping ${\Gamma}$ from $\mathbf{q}$ to the steady-state market sizes $(R_0, D_0)$ is continuous. 
    
    Although these formulas may return a negative idle-driver pool for some assumed profiles, we extend the simulation mapping to all real-valued inputs by setting
\[ \Psi(R,D)=\mathbf{e}_0 \qquad\text{whenever }R\leq 0\text{ or }D\leq 0. \]
    
    For the empirical mapping ${\Psi}$, non-integer values of expected market sizes $R_0$ and $D_0$ must be passed into a discrete-event simulator. We handle this via probabilistic rounding: a continuous value $x$ is instantiated as $\lceil x \rceil$ with probability $x - \lfloor x \rfloor$, and as $\lfloor x \rfloor$ with probability ${1-(x - \lfloor x \rfloor)}$. This probabilistic interpolation ensures that the expected input distribution to the simulation environment varies continuously with the theoretical market thickness $(R_0, D_0)$. Consequently, the expected empirical output $\mathbf{q}'$ also varies continuously with respect to the input $(R_0, D_0)$. 

    Because both ${\Gamma}$ and ${\Psi}$ are continuous, their composition ${\Phi = \Psi \circ \Gamma}$ is continuous over the domain. Since ${\Phi}$ is a continuous mapping from a compact, convex set to itself, Brouwer's Fixed Point Theorem guarantees the existence of at least one fixed point $\mathbf{q}^* \in \Delta^U$ satisfying ${\Phi}(\mathbf{q}^*) = \mathbf{q}^*$.

    Finally, suppose that $R_0(\mathbf{q}^*)\leq 0$ or $D_0(\mathbf{q}^*)\leq 0$. By the definition of ${\Psi}$ and the fixed-point identity, this would imply $\mathbf{q}^*=\mathbf{e}_0$. However, the flow balance equations give
\[ \Gamma(\mathbf{e}_0) = \left(\frac{\lambda}{\eta}, \frac{\widetilde{\lambda}}{\widetilde{\eta}_I}\right), \]
whose components are strictly positive. This is a contradiction. Hence, the fixed point has strictly positive market sizes.
\end{proofof}
}

Finally, we utilize the matrix representation of the system dynamics to derive closed-form expressions for the match probability and the expected match latency. Let $M$ be the $m \times m$ transient rate matrix for the rider states, where $m=U+1$ for FA and $m=2U$ for BA. Furthermore, let $Q$ be the $m \times 2$ absorption rate matrix, where the first column corresponds to reneging (all entries equal to $\rideChurn$) and the second column corresponds to the ``Matched'' state. The specific entries are:
\begin{equation*}
\textbf{FA:}\quad Q_{i2} = (i-1)\mu p\qquad,\qquad
    \textbf{BA:}\quad Q_{i2} = \mu p~\textrm{for}~1<i < 2U~\textrm{and}~Q_{(2U)2} = \mu + \notifChurn.
\end{equation*}
Using this framework, we derive the match probability and match time in the following proposition.

\begin{proposition} \label{prop:MatchTime}
    Define the fundamental matrix as $N = -M^{-1}$. For a rider starting in state $i$, the probability of being matched and the expected time to a successful match are given by:
    \begin{align*}
        \prob\left[\text{Matched} \mid X_0 = i \right] &= \left(N Q \right)_{i 2}\qquad , \qquad \mathbf{E}\left[\,\text{Match Time} \;\middle|\; X_0 = i,\, \text{Matched} \,\right] = \left(N^2 Q \right)_{i 2}/ \left(N Q \right)_{i 2}.
    \end{align*}
\end{proposition}
\vspace{-2mm}

\subsection{Stylized Simulation Setup to Analyze Algorithmic Equilibrium}\label{Sec:MMMSimulation}
\begin{figure}[htb]
    \centering
    \includegraphics[width=0.5\linewidth]{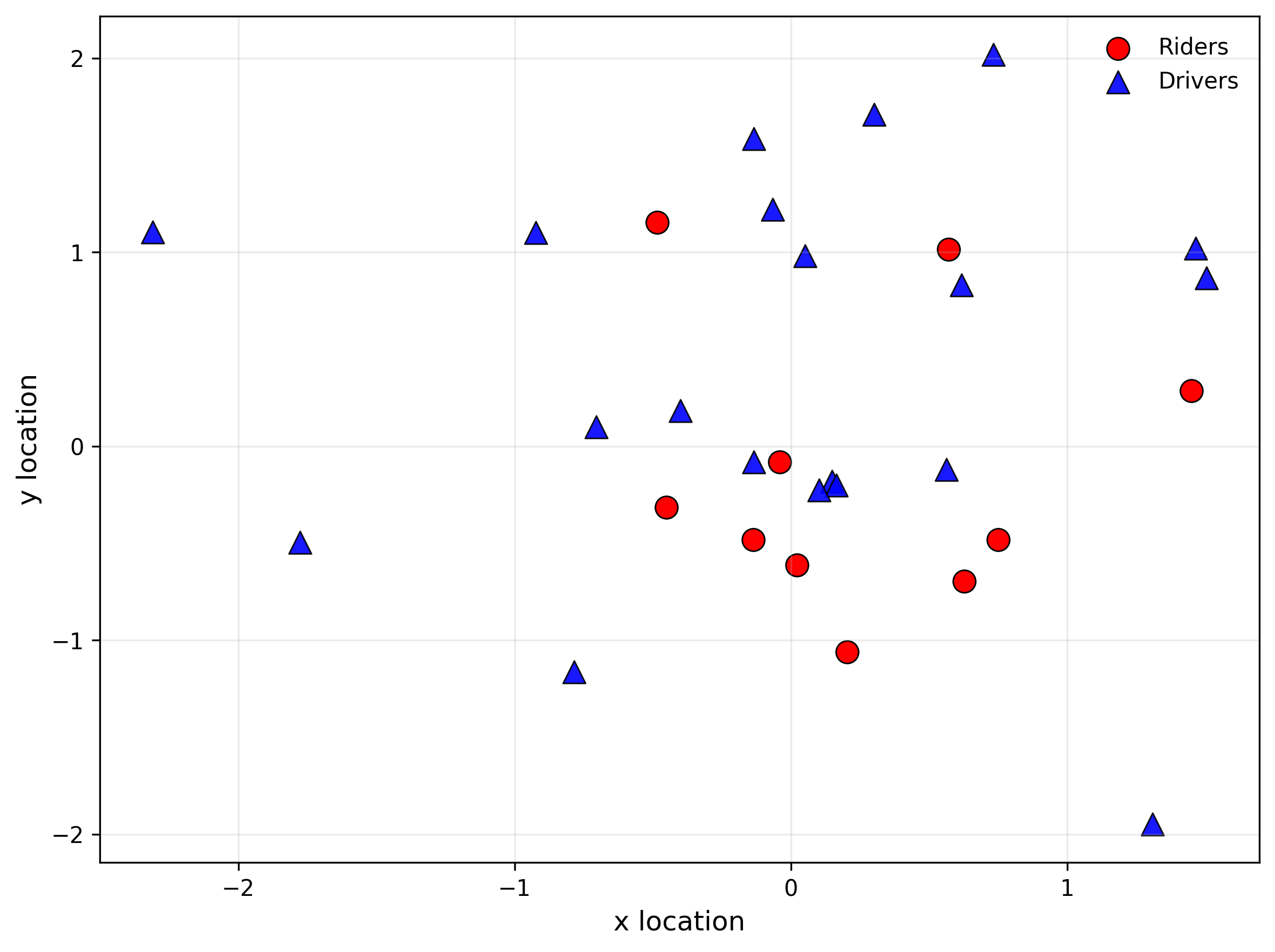}
    \caption{Spatial distribution of 10 riders and 20 drivers in the stylized simulation setup.}
    \label{fig:spatialDistribution}
\end{figure}

We consider a marketplace populated by $R_0$ riders and $D_0$ drivers. We make the assumption that these agents are geographically distributed in a two-dimensional Euclidean plane. The coordinates $(x, y)$ of all agents are drawn independently from a standard normal distribution $\mathcal{N}(0, 1)$, reflecting the clustering of demand and supply in urban centers (\Cref{fig:spatialDistribution}). The utility of a match is modeled as a decaying function of distance. Specifically, the score $w_{r,d}$ for a match between a driver and rider separated by Euclidean distance $distance$ is defined as $w_{r,d} = \frac{1}{1+distance}$.\footnote{Similar to \Cref{sec:simulation}, this stylized choice is due to the fact that we cannot use Lyft's original scoring system in a simulation with counterfactual pairs, due to certain technical limitations beyond the scope of this work.} 

Given any dispatch policy (notification packing and contention resolution) and fixing the market size parameters $(R_0,D_0)$, we first define a mapping from $(R_0,D_0)$ to a probability distribution over notification sizes, denoted by the vector $\mathbf{q} = [q_1, \dots, q_U]$ (as described \Cref{sec:equilibrium_method}; see also Figure~\ref{fig:stylized_model_cycle}). This mapping is implicitly constructed via running a simulation in the above stylized marketplace . \revcolor{More specifically, we focus on  the algorithms we discussed in \Cref{sec:Algs}---including \textsc{Opt}, \textsc{Approx}, \textsc{Greedy}, and \textsc{Ed+}\footnote{Note that \textsc{Rejection-aware} will be the same as ED as $p > 0.2$.}---under contention resolution protocols FA and BA.} For each of these dispatch procedures, given input market size $(R_0,D_0$), we run a Monte-Carlo simulation, where in each iteration we sample the spatial locations of $R_0$ riders and $D_0$ drivers to create a static snapshot of the market. Based on these locations and hence the scoring function, we simulate the policy for a single cycle using the framework in \Cref{sec:simulation}, and take the average over samples to obtain an estimated distribution over notification sizes $\mathbf{q}$.

\begin{figure}[htb]
    \centering
    \includegraphics[width=0.65\linewidth]{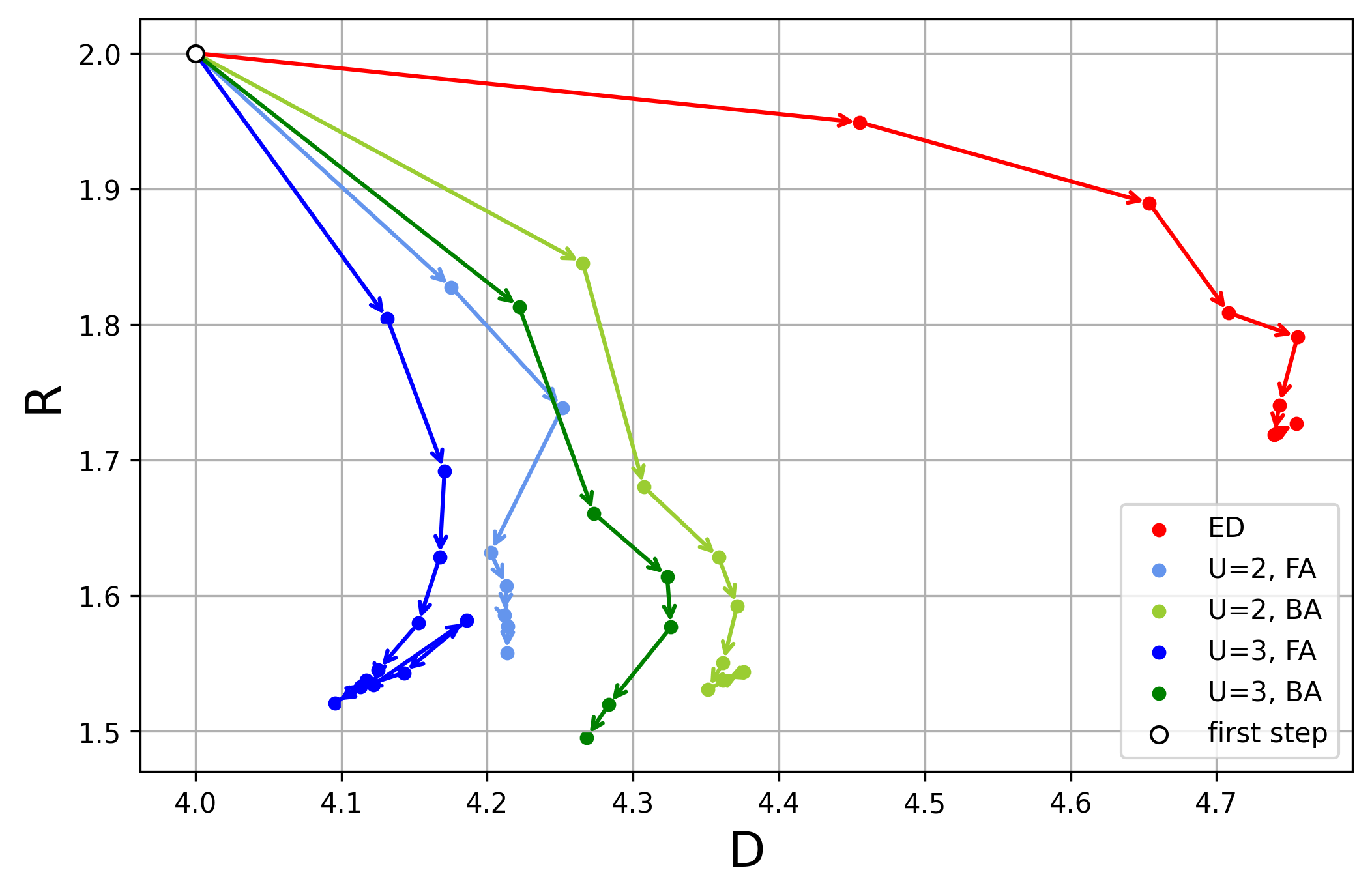}
    \caption{Convergence trajectory of \textsc{Opt} policies and ED to their respective equilibria. Starting from a common initial state, the paths trace the iterative evolution of $(R_0,D_0)$ market sizes under different algorithmic configurations ($U \in \{2, 3\}$, FA vs.\ BA). The markers denote the state at each iteration of the fixed-point computation.}
    \label{convergence-trajectory}
\end{figure}

Having query access to the above mapping, we plug it into our iterative fixed-point approach described in \Cref{sec:equilibrium_method}, which in turn uses the theoretical fixed-point equations established in \Cref{sec:MMMFA} and \Cref{sec:MMMBA} as explained earlier. This allows us to calculate the resulting long-term steady-state market thickness $(R_0, D_0)$ for that specific dispatch policy, and to compute the induced equilibrium outcomes. Somewhat interestingly, we observe quite fast convergence in our simulations. For example, see Figure~\ref{convergence-trajectory} for the convergence trajectories of optimal NED and ED under FA/BA to their respective equilibria.

\smallskip
\xhdr{The impact of algorithmic choices on market equilibrium.} Changing the core notification packing algorithm of the dispatch policy itself (alongside its key parameters, such as the maximum notification set size $U$ and the optimization threshold $\theta$), and, critically, the contention resolution protocol, fundamentally alters the notification distribution $\mathbf{q}$. As shown in Figure~\ref{fig:RD_MMM_scatter}, this ultimately drives a shift in the market equilibrium---i.e., the equilibrium size/thickness of the market on both riders and drivers side, specified by parameters $(R_0,D_0)$ as detailed earlier.

A notable managerial observation concerns the equilibrium driver count under the BA protocol. Intuitively, since the BA notification process is sequential and thus slower per request, one might expect a larger backlog of busy drivers, resulting in lower availability ($D_0$) compared to FA. However, our results show the opposite. The FA protocol achieves a higher match rate, meaning it successfully transitions more drivers from the ``idle'' state to the ``en-route/busy'' state. In the regime we studied, this higher throughput dominates the latency effect: FA depletes the idle driver pool more aggressively than BA, resulting in a lower equilibrium $D_0$.

\smallskip
\xhdr{Performance landscape.} \revcolor{ \Cref{fig:RD_MMM_scatter} illustrates the performance trade-offs associated with various policies we described in \Cref{sec:Algs} across the ($U, \theta$) parameter space. Refer to \Cref{sec:MMM-sensitivity} for a sensitivity analysis with respect to these parameters. It is important to note the hybrid methodology used to generate these insights: the ``Score'' is a  simulation-based \revcolor{estimate} derived from the spatial utility of matches in the equilibrium, whereas the ``Match Probability'' and ``Match Time'' are theoretical values computed by evaluating the steady-state expressions from \Cref{prop:MatchTime} at the induced equilibrium.

\begin{figure}[htb]
    \centering
        \includegraphics[width=\linewidth]{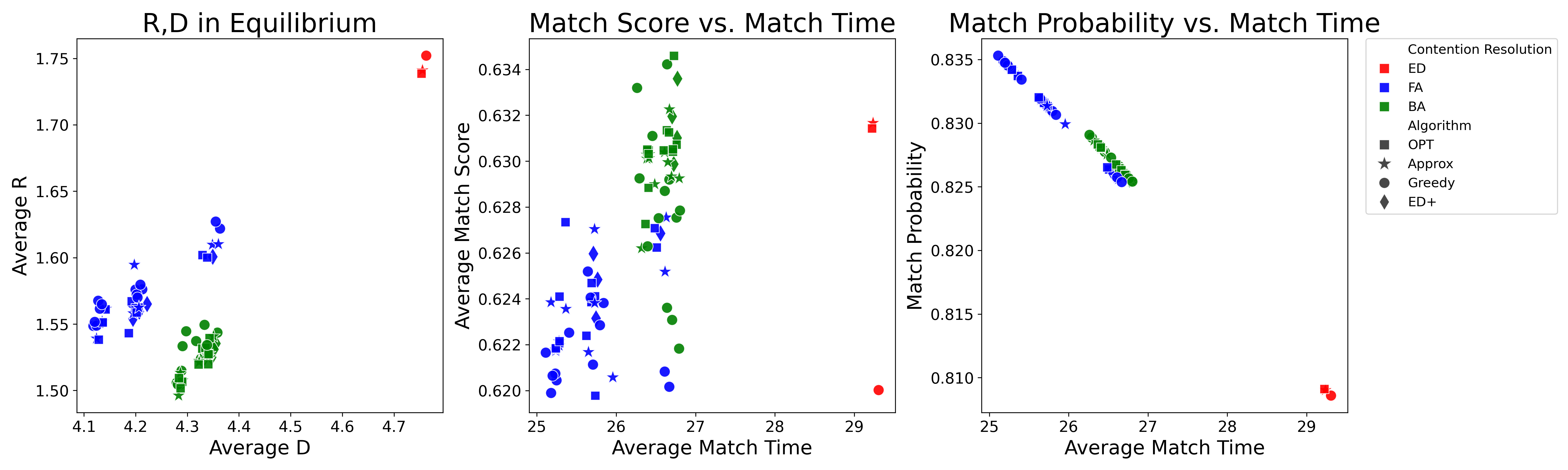}
    \caption{Equilibrium outcome of \{\textsc{Opt}, \textsc{Approx}, \textsc{Greedy}, \textsc{Ed+}\} under different parameters and contention-resolution protocols. Parameters: $\notifChurn = 0, \idleChurn = \rideChurn = 0.02$, $p = 0.4$, $\mu=1/3$, and $\riderTotalArrival = \driverArrival = 1/2$.}
    \label{fig:RD_MMM_scatter}
\end{figure}

The equilibrium results show that exclusive dispatch achieves a slightly higher match score than many NED algorithms in the stylized model. This differs from the trends in \Cref{fig:scatter} because all drivers now share the same acceptance probability. This simplifying assumption removes the variation in acceptance probabilities that NED can exploit when choosing notification sets. The score advantage of exclusive dispatch is modest, however, at approximately 1--2\%, while NED reduces match time by approximately 10\%.

The equilibrium results also preserve the operational trade-off between contention-resolution protocols. BA achieves higher match quality, whereas FA provides faster matching and greater matching throughput. Thus, accounting for long-term market dynamics continues to favor BA when match quality is the priority and FA when service speed and throughput are more important.}

\smallskip
\xhdr{Model limitations.} \revcolor{We note a structural limitation of this stylized model compared to the high-fidelity simulation in \Cref{sec:simulation}. To evaluate system performance, we first determine the equilibrium market size using the iterative procedure outlined in \Cref{fig:stylized_model_cycle}. Once the steady-state pool size is established, we compute the average match score by halting new arrivals and running the algorithms on the resulting fixed graph—accounting for potential reneging—until all viable edges have been evaluated. We explicitly design this exhaustive, multi-cycle approach to avoid the severe optimistic bias inherent in a single-cycle evaluation. Because our macroscopic model samples a fresh graph for each evaluation, calculating a single-cycle average score would naively assume that when algorithms like ED successfully make a match, they always pair riders with their absolute closest drivers on the first attempt. By forcing the algorithms to process rejections and secondary matches on a static graph until no edges remain, we attempt to strip away this unrealistic upper bound and obtain a much fairer approximation of true algorithmic performance. Yet, this approach is still not ideal.

Furthermore, while the macroscopic model effectively captures steady-state pool sizes for both riders and drivers, it relies on independent draws between samples. Whether this abstraction disproportionately favors NED or ED remains ambiguous, for two reasons. First, in reality, ED operates sequentially: if a primary driver rejects a dispatch, that driver remains in the system and may still be available for subsequent requests. The stylized model's assumption of independent draws abstracts away this temporal persistence, which can bias the estimated efficiency of ED in either direction. Second, the model assumes that a larger, ``thicker'' market uniformly improves match probability. In practice, however, part of the excess market thickness often manifests as an accumulation of drivers and riders that were difficult to match in previous cycles—for instance, drivers located at the spatial peripheries of a city where matches are inherently more difficult to form. Because ED tends to maintain a thicker overall market due to its slower matching dynamics, our macroscopic model likely overvalues the practical utility of ED's larger pool size, which will be more skewed toward the borders of the city. Consequently, the net direction of the bias introduced by these spatial and temporal abstractions is ambiguous: the first can bias the comparison in either direction, whereas the second likely favors ED}.

\section{Conclusions \& Future Directions}
\label{sec:conclusion}
In this work---part of our collaboration with Lyft and paired with our companion paper~\ifblind\cite{LyftI2026companion_Blind}\else \cite{LyftI2026companion_NonBlind}\fi---we study how dispatch policies shape long-run performance in ride-hailing. We combine two complementary approaches---first an empirical approach based on a large-scale simulations driven by Lyft data and second a theoretical approach based on a stylized Markovian macro model that isolates equilibrium forces. This combination allowed us to validate our conclusions in both complex, real-world scenarios and controlled theoretical environments.


Our analysis yields several key takeaways, and we highlight some of them here (see the full list with more details in \Cref{sec:simulation} and \Cref{sec:MMM}). First, non-exclusive dispatch (NED) consistently outperforms traditional exclusive dispatch (ED) across a wide range of market conditions. Crucially, this advantage remains robust even when the platform has only coarse estimates of driver acceptance probabilities, indicating that the benefits of NED are not contingent on perfect behavioral estimation. Second, we quantify a fundamental trade-off between contention-resolution protocols: First-Accept (FA) maximizes speed and liquidity, whereas Best-Accept (BA) is required to maximize per-match quality. Third, both the equilibrium analysis and the simulations show sharply diminishing returns to increasing the notification bound $U$, suggesting that small notification sets can capture most of NED's value while limiting communication overhead.


Several avenues for future research remain. First, while we modeled driver acceptance as stochastic, future work should consider strategic behavior, where drivers might gamify their response times under different protocols. Second, our findings suggest the potential for adaptive policies that dynamically toggle between FA and BA or adjust notification bounds based on real-time market states. Finally, developing simple, interpretable packing heuristics (or algorithms with theoretical performance guarantees) that capture the core dynamic effects remains an interesting direction.



\newcommand{\newblock}{}
\setlength{\bibsep}{0.0pt}
\bibliographystyle{plainnat}
{\footnotesize
\bibliography{refs}}

\renewcommand{\theHchapter}{A\arabic{chapter}}
\renewcommand{\theHsection}{A\arabic{section}}

\newpage
\clearpage
\normalsize
\pagestyle{ECheadings}%
\ECHowEquations
\ECHowSections
\setcounter{figure}{0}%
\renewcommand{\thefigure}{EC.\arabic{figure}}%
\setcounter{table}{0}%
\renewcommand\thetable{EC.\@arabic\c@table}%
\setcounter{page}{1}\def\thepage{ec\arabic{page}}%
\hspace*{1em}

\ECDisclaimer

\section{Related Work}\label{app:related-work}
This appendix provides a more detailed discussion of how our work relates to existing research in operations research, economics, and computer science.

\smallskip
\xhdr{Dynamic matching and matching queues.}
A parallel stream of work analyzes dispatch as a \emph{dynamic matching} or \emph{matching-queue} problem in transportation and ride-hailing platforms. Classical and recent models study steady-state performance and design of matching policies under stochastic arrivals, waiting costs, abandonment, and post-allocation services; see, e.g., \cite{ozkan2020dynamic,aouad2020dynamic,ashlagi2019edge,amanihamedani2024improved,aveklouris2025matching,akbarpour2014dynamic,ata2020dynamic,bansak2024dynamic} and references therein. These works typically focus on designing dynamic policies and analyzing steady-state performance under binding allocations. Our paper shares their long-run view but differs structurally: driver participation is explicitly stochastic at the dispatch stage, and the platform can issue multiple notifications for the same ride. Our aim is not to propose an optimal policy for a stylized queueing model, but to quantify and explain the long-run trade-offs induced by simple, implementable design choices.

Closest in spirit in the above literature are models that treat offers as non-binding and incorporate reneging and acceptance frictions. For instance, \citet{castro2020matchingqueues} analyze matching queues with reneging; \citet{castro2022randomized} propose randomized FIFO-style mechanisms for sequential offer processes; and \citet{castro2020matching} study flexibility incentives in matching queues. Related models also incorporate strategic and preference-driven acceptance (with abandonment and randomized priority rules), such as strategic servers in ride-hailing \cite{varma2021near} and driver location preferences \cite{rheingans2019ridesharing}. While these papers share with us the emphasis on long-run marketplace effects, our setting differs from these sequential-offer models by emphasizing \emph{simultaneous} broadcast notifications in each cycle and by explicitly comparing contention-resolution protocols that map multiple acceptances to a single match.

\smallskip
\xhdr{Non-exclusive dispatch, broadcasting, and order-grabbing in ride-hailing.} A growing transportation science literature studies variants of non-exclusive (broadcast) dispatch in ride-hailing and taxi-hailing systems, in which requests are \emph{broadcast} or ``informed'' to multiple drivers without immediate commitment, and drivers decide whether to \emph{grab} the order. For example,
on the modeling side, \citet{sun2020taxi} study whether a platform should \emph{inform} multiple drivers or \emph{assign} a ride to a single driver, and study their operational implications. More recent work in transportation literature study richer broadcast and multi-round procedures in which a request can be offered to multiple drivers and refined over time; for example, \citet{qin2025two_round} propose a two-round broadcasting mechanism to reduce idle driving time, and \citet{feng2022block} study multi-driver dispatch via block matching. Other empirical and learning-oriented work studies how the matching radius and driver response behavior interact under broadcasting, including dynamic matching radii and order-grabbing behavior \cite{chen2025dynamic,chen2025grab,chen2023dynamic}.

These papers share with our setting the key feature that notifications are \emph{non-binding} and acceptance behavior is uncertain. Our focus differs in two ways. First, we explicitly model the contention-resolution step that follows broadcast notifications and compare two canonical protocols used in practice (First-Accept versus Best-Accept). Second, we study long-run marketplace effects: we combine a high-fidelity discrete-event simulator calibrated to Lyft data with a stylized macro model that explains why the FA--BA trade-off persists at equilibrium.

\smallskip
\xhdr{Matching and pricing at ride-hailing platforms.}
Our study is motivated by operational questions at Lyft and complements a growing body of research on marketplace mechanisms developed for ride-hailing platforms to handle high-volume dispatch. Recent work analyzes design trade-offs such as the Pareto frontier between shared-ride value and detours \cite{lobel2025detours}, the adoption benefits of static versus dynamic pickup locations~\cite{yan2025trading}, the effect of batching in matching and pricing~\cite{feng2024two,feng2025batching}, and the use of reinforcement learning to move from greedy heuristics to global value maximization \cite{azagirre2024better}. More broadly, the systems and ML communities have proposed large-scale dispatch and repositioning methods for ride-hailing platforms, including learning-and-planning dispatch pipelines \cite{xu2018largeScaleDispatch}, mean-field and multi-agent reinforcement learning for dispatch and matching \cite{qin2019meanFieldRidesharing,qin2020didiDispatchRL,ke2019multiAgentDispatch}, and reinforcement-learning repositioning policies \cite{jiao2021repositioningRL}.  Complementary OR research studies matching and pricing interactions \cite{yan2020dynamicPricingMatching,garg2022driverSurge,castillo2025matching}. Parallel streams of research---often in collaboration with ride-hailing services and Autonomous Vehicle (AV) platforms---study the economic and operational integration of mixed fleets; see, for instance, \cite{castro2024autonomous,freund2025supply,lian2025capturing}. 
Our work complements this literature by isolating the \emph{notification layer} of the stack its long-run implications when acceptance is imperfect. We analyze how the choice of packing algorithm and contention-resolution protocol (FA vs.\ BA) determines whether downstream objectives can be realized in a stochastic environment characterized by imperfect driver acceptance behavior. 

\revcolor{More related to our marketplace macro model, \cite{wang2024demand} study ride-hailing dynamics through a spatial queueing model in which pickup times depend on market thickness, and use a fluid approximation to characterize the long-run equilibrium. Our setting is more complex because drivers may reject notifications and because non-exclusive dispatch allows multiple drivers to be notified simultaneously under different contention-resolution protocols. This significantly enlarges the state space and makes it difficult to adapt a stylized theoretical model of pickup times without imposing additional assumptions. We therefore use Monte Carlo simulations based on real Lyft data to capture these pickup-time dynamics more faithfully.}

\smallskip
\xhdr{Supplier menus and choice-based dispatch.}
Another related thread studies \emph{menus} in dynamic matching markets, where the platform presents a set of options to a strategic supplier (e.g., a driver) who then chooses an assignment. In the ride-sourcing and logistics context, \citet{ausseil2022supplier} formulate the platform problem as designing supplier menus that maximize expected performance under stochastic choice. Hierarchical and assortment-style menu designs have also been proposed in peer-to-peer logistics and on-demand matching \cite{mofidi2019beneficial,horner2021optimizing,horner2025increasing,yang2022menus}. While our NED mechanism presents at most one offer to each driver (to reflect practical UI constraints), the menu literature is conceptually aligned in that it gives more options to drivers. Also, compared to our setting, these models often place additional emphasis on strategic choice and incentive effects of the menu design. Our focus is different: \RNcolor{we treat acceptance behavior as an exogenous stochastic input that comes from estimations (and also consider homogeneous-acceptance policies for fairness and practicality that use the same global estimate)}.

\smallskip
\xhdr{FA/BA-style broadcast allocation beyond ride-hailing.}
Broadcast-offer allocation with stochastic participation---where allocation goes to the first responder (FA) or to the best among (timely) responders (BA)---also appears in several other application domains. Examples include food rescue and donation platforms \cite{alptekinoglu2024achieving,lee2025offer,shi2020improving}, community first-responder dispatch \cite{henderson2022should,DellaertSchlicherHillenaarJagtenberg2024CFRNetworks}, and spatial crowdsourcing systems that multicast tasks and then finalize among responders \cite{basik2018fair}. Another possible application is deceased-donor transplantation, where organ offers are extended to candidates and centers whose acceptance behavior is heterogeneous and time-sensitive \cite{wey2017influence,husain2019association,agarwal2025choices}. A closely related BA instantiation arises in online advertising ``header bidding,'' where impressions are broadcast to multiple exchanges and the highest timely bid wins \cite{pachilakis2019no,aqeel2020untangling}. 

Closest to us in terms of the model and technical framework is the concurrent work by \cite{liu2025recommend} that studies a two-stage ``recommend-to-match'' problem for crowd-sourcing logistics/freight platforms under stochastic supplier rejections. Their formulation coincides with the Best-Accept objective in our model (up to an additional constraint that each request is recommended to at most some number of suppliers) and enforces the same supply-side exclusivity (each supplier receives at most one recommendation). The focus of this paper is different from ours, as they consider tractable mathematical programming approaches, giving an exact MILP for the homogeneous-acceptance special case and proposing a mixed-integer exponential cone approximation with parametric performance bounds and extensive numerical evaluation. Our companion paper \ifblind\cite{LyftI2026companion_Blind}\else \cite{LyftI2026companion_NonBlind} \fi looks at a similar algorithmic problem. In contrast to this paper, we provide an \emph{exact polynomial-time} algorithm for the BA homogeneous-acceptance special case in this companion work. Also, our companion work provides complexity results and worst-case approximation algorithms for both Best-Accept and First-Accept contention protocols in the general case, including a PTAS/constant-factor guarantees and improved approximation beyond $1-1/e$ for BA via demand-oracle methods.

\section{Missing Details of \Cref{sec:simulation}}
\label{apx:simulations}

\subsection{NED Packing Algorithms}
\label{sec:Algs}
\revcolor{Throughout our simulations, we evaluate the performance of several distinct algorithmic strategies.

\subsubsection{\textsc{Opt} (Optimal packing)}\label{sec:optALG}
The first approach we consider involves solving the optimization problem defined in \eqref{prob:packing} to optimality. As previously noted, while this problem is computationally demanding in the general case, the sparsity inherent in practical ride-hailing graphs (as detailed in \Cref{sec:Data}) renders it solvable within seconds for typical instances. Efficient implementation also requires specific data preprocessing steps.

This algorithm is governed by two key parameters: the maximum notification batch size $U$ and the opportunity cost threshold $\theta$. Varying these parameters allows us to modulate the system's behavior along the spectrum of temporal foresight. Specifically, large values of $U$ combined with low values of $\theta$ induce a more \emph{myopic} policy, prioritizing immediate match probability over the conservation of driver supply for future intervals. A particularly significant special case arises when $U=1$ and $\theta=0$. Under this parameterization, the formulation reduces to a standard Maximum Weight Matching problem. This configuration effectively models the classical exclusive dispatch system (ED), serving as a baseline to quantify the specific efficiency gains derived from NED notifications.

\subsubsection{\textsc{Greedy} (Algorithm~\ref{alg:greedy})}
Motivated by the greedy algorithm for submodular welfare maximization, we employ a sequential heuristic to solve the packing problem. This approach iterates through the available drivers and tentatively assigns each to the ride that yields the highest marginal increase in expected welfare, provided the notification batch size does not exceed $U$. To ensure efficiency, the assignment {of driver $d$ to ride $r$} is finalized only if the marginal gain outweighs the weighted opportunity cost {$\theta p_{r,d}$, as in the marginal efficiency constraint of \eqref{prob:packing}}.

\smallskip
\begin{algorithm}[H]
\caption{\textsc{Greedy} Algorithm}\label{alg:greedy}
\textbf{Input:} Set of drivers $D$, set of rides $R$, threshold $\theta$, batch limit $U$
\For{ride $r \in R$}{
    Initialize notification list $\mathcal{N}_r \gets \emptyset$\;
}
Fix a random ordering of drivers: $\tilde{D} = [d_1, d_2, \ldots, d_n]$\;
\For{driver $d \in \tilde{D}$}{
    \For{ride $r \in R$ with $|\mathcal{N}_{r}| < U$}{
        Compute marginal gain: $\Delta_{d,r} = F_r(\mathcal{N}_r \cup \{d\}) - F_r(\mathcal{N}_r)$\;
    }
    Let $r^* = \arg\max_{r \in R} \Delta_{d,r}$\;
    \If{$\Delta_{d,r^*} > \theta {p_{r^*,d}}$}{
        $\mathcal{N}_{r^*} \gets \mathcal{N}_{r^*} \cup \{d\}$\;
    }
}
\end{algorithm}

\subsubsection{\textsc{Rejection-aware} (Algorithm~\ref{alg:rej_aware})}
We evaluate another heuristic designed to approximate practical dispatch logic. Unlike the previous approach, which processes drivers sequentially, this algorithm processes rides sequentially. For each request, it greedily constructs a notification set by selecting the highest-quality available drivers. This aggregation continues until a reliability condition is met, specifically, until a driver with a sufficiently low rejection probability (e.g., $< 0.8$) is included, or until the batch reaches a hard cap of $U$. Any agents remaining after this sequential pass are matched using a standard one-to-one Maximum Weight Matching algorithm.

The rationale behind this heuristic is to prevent over-notification, that is, notifying a large set of drivers for the same ride, for easy-to-match rides. By ensuring that a request does not unnecessarily tie up multiple drivers with high acceptance likelihoods, the algorithm reduces the inefficiency associated with resolving contention among multiple accepting drivers.

\smallskip
\begin{algorithm}[H]
\caption{\textsc{Rejection-aware} Algorithm}\label{alg:rej_aware}
\textbf{Input:} Set of drivers $D$, set of rides $R$, rejection likelihood function $\text{Rej}(\cdot)$, batch limit $U$

\For{ride $r \in R$}{
    Initialize notification list $\mathcal{N}_r \gets \emptyset$\;
}

Fix a random ordering of rides: $\tilde{R} = [r_1, r_2, \ldots, r_n]$\;

\For{ride $r \in \tilde{R}$}{
    Let $\tilde{D}_r = [d_{r,1},\ldots,d_{r,n_r}]$ be the list of available drivers in $D$, sorted in decreasing order of weight ${w_{r,d}}$\;
    
    $\tilde{\mathcal{N}} \gets \emptyset$\;
    
    \For{driver $d \in \tilde{D}_r$}{
        $\tilde{\mathcal{N}} \gets \tilde{\mathcal{N}} \cup \{d\}$\;
        
        \If{$\text{Rej}(d) < 0.8$ \textbf{or} $|\tilde{\mathcal{N}}| \geq U$}{
            break\;
        }
    }
    
    \If{$|\tilde{\mathcal{N}}| > 1$}{
        $\mathcal{N}_{r} \gets \tilde{\mathcal{N}}$, $D \gets D \setminus \tilde{\mathcal{N}}$, 
        $R \gets R \setminus \{r\}$\;
    }
}
Find the maximum weight matching \FEedit{(ED)} between remaining $D$ and $R$.
\end{algorithm}

\subsubsection{\textsc{Ed+} (Algorithm~\ref{alg:enhanced_nd})}
We also consider the \FEedit{Enhanced Exclusive Dispatch (\textsc{Ed+}) (Algorithm~\ref{alg:enhanced_nd})} algorithm, a hybrid strategy that improves upon the standard ED baseline. The algorithm operates in two distinct phases. First, it computes a global Maximum Weight Matching (the standard ED solution) to assign one available driver $d_r^*$ to ride $r$ in a one-to-one fashion. In the second phase, the algorithm attempts to enhance these assignments by identifying a second-best driver from the \emph{remaining} unassigned pool. If adding this secondary driver{, $d$,} increases the expected welfare by more than the threshold {$\theta p_{r,d}$}, the notification set is expanded; otherwise, the original single assignment is preserved. This ensures that the system retains the efficiency of the standard global matching while selectively reinforcing specific high-value or high-risk requests.

\smallskip
\begin{algorithm}[h!]
\caption{\textsc{Ed+} Algorithm}\label{alg:enhanced_nd}
\textbf{Input:} Set of drivers $D$, set of rides $R$, threshold $\theta$

Compute the maximum weight matching $\mathcal{M}$ between $D$ and $R$. Let $(r, d_r^*) \in \mathcal{M}$ denote the matched pairs.

Update available drivers: $D \gets D \setminus \{d_r^* \mid (r, d_r^*) \in \mathcal{M}\}$\;

Fix a random ordering of matched rides: $\tilde{R} = [r_1, r_2, \ldots, r_k]$\;

\For{ride $r \in \tilde{R}$}{
    Initialize notification list $\mathcal{N}_r \gets \{d_r^*\}$\;

    Let $d^{\diamond}_r = \arg\max_{d \in D} {w_{r,d}}$\;

    \If{$d^{\diamond}_r \neq \emptyset$ \textbf{and} $F_r(\{d_r^*,d^{\diamond}_r\}) \geq F_r(\{d_r^*\}) + \theta {p_{r,d^{\diamond}_r}}$}{
            $\mathcal{N}_{r} \gets \{d_r^*,d^{\diamond}_r\}$\;
            
            $D \gets D \setminus \{d^{\diamond}_r\}$\;
    }
}
\end{algorithm}

\subsubsection{\textsc{Approx} (Algorithms~\ref{alg:multi} and \ref{alg:derandomized})}
\textsc{Approx} is the LP-based algorithm of \ifblind\cite{LyftI2026companion_Blind}\else\cite{LyftI2026companion_NonBlind}\fi. In each cycle it (i) solves a linear program that is the relaxation of the notification problem, (ii) independently rounds the fractional solution into disjoint notification sets, and (iii) trims each set to respect the batch-size limit $U$ and the opportunity-cost threshold $\theta$. We describe each step below.

\paragraph{Step 1: LP relaxation.}
For each rider--driver pair $(r,d)$, let $x_{r,d}$ denote the probability that driver $d$ is notified about ride $r$, and let $y_{r,d}$ denote the probability that the pair is ultimately matched. The LP relaxation is

\begin{align}
\label{eq:LP}\tag{LP}
\max_{x,y\geq0}\quad
& \sum_{r\in\mathcal{R}}\sum_{d\in\mathcal{D}}
w_{r,d}y_{r,d}\\
\text{s.t.}\quad
& y_{r,d}\leq p_{r,d}x_{r,d}
&& \forall r\in\mathcal{R},\ d\in\mathcal{D}, \nonumber\\
& \sum_{r\in\mathcal{R}}x_{r,d}\leq1
&& \forall d\in\mathcal{D}, \nonumber\\
& \sum_{d\in\mathcal{D}}y_{r,d}\leq1
&& \forall r\in\mathcal{R}. \nonumber
\end{align}
The first constraint states that a match requires the driver to be notified and to accept. The second states that each driver receives at most one notification. The third states that each ride is matched to at most one driver. The LP does not model contention among the drivers notified for the same ride, so its solution is fractional and generally cannot be implemented directly as notification sets.

\paragraph{Step 2: Rounding.}
Algorithm~\ref{alg:multi} turns an optimal LP solution $(x^*,y^*)$ into disjoint notification sets. Each driver $d$ is independently assigned to at most one ride: ride $r$ is selected with probability $y^*_{r,d}/p_{r,d}$, and the driver is left unassigned with the remaining probability. If ride $r$ is selected, the notification is retained with probability $1-y^*_{r,d}/3$. This attenuation step thins out the pairs with large matching marginals, which would otherwise crowd out the other drivers notified for the same ride under FA.

\begin{algorithm}[H]
\caption{Independent rounding with attenuation}
\label{alg:multi}
\KwIn{Weights $\{w_{r,d}\}$ and acceptance probabilities $\{p_{r,d}\}$}
\KwOut{Disjoint notification sets $(S_r)_{r\in\mathcal{R}}$}

Solve \eqref{eq:LP} to obtain an optimal solution $(x^*,y^*)$\;
Set $S_r\gets\varnothing$ for every $r\in\mathcal{R}$\;
\ForEach{$d\in\mathcal{D}$}{
    Independently sample $R_d\in\mathcal{R}\cup\{\varnothing\}$ with
    \[
    \Pr(R_d=r)=
    \begin{cases}
        y^*_{r,d}/p_{r,d}, & p_{r,d}>0,\\
        0, & p_{r,d}=0,
    \end{cases}
    \qquad
    \Pr(R_d=\varnothing)
    =1-\sum_{r:p_{r,d}>0}\frac{y^*_{r,d}}{p_{r,d}}
    \]
    \If{$R_d=r\neq\varnothing$}{
        Independently retain the notification with probability
        $1-y^*_{r,d}/3$\;
        If retained, set $S_r\gets S_r\cup\{d\}$\;
    }
}
\KwRet{$(S_r)_{r\in\mathcal{R}}$}\;
\end{algorithm}
\paragraph{Derandomization.}
In our simulations we replace the random choices of Algorithm~\ref{alg:multi} with the deterministic procedure in Algorithm~\ref{alg:derandomized}, so that each cycle returns a single set of notifications. The procedure fixes the drivers one at a time (in a fixed order) while keeping the not-yet-processed drivers random. To evaluate a partially fixed solution, we track for each ride $r$ and driver $d$ the \emph{effective acceptance probability} $u_{r,d}$, i.e., the probability that $d$ is notified about $r$ and accepts. Given $\boldsymbol{u}_r=(u_{r,d})_{d\in\mathcal{D}}$, the expected FA value of ride $r$ is
\[
F_r^{\text{FA}}(\boldsymbol{u}_r)
\;=\;
\sum_{d\in\mathcal{D}} w_{r,d}\,u_{r,d}\int_0^1\prod_{\ell\in\mathcal{D}\setminus\{d\}}\bigl(1-u_{r,\ell}\,t\bigr)\,dt ,
\]
which is the FA valuation with the acceptance probabilities replaced by $\boldsymbol{u}_r$. The vector $\boldsymbol{u}$ is initialized to the values induced by the randomized rounding, $u_{r,d}=y^*_{r,d}(1-y^*_{r,d}/3)$. When driver $d$ is processed, we compare (a) leaving $d$ unnotified, which sets $u_{r,d}=0$ for all $r$, with (b) notifying $d$ about ride $r$, which sets $u_{r,d}=p_{r,d}$ and keeps $u_{r',d}=0$ for $r'\neq r$. We pick the ride with the largest gain over option (a), and notify $d$ only if this gain is strictly positive. We write $\boldsymbol{u}_r^{\,d\leftarrow p_{r,d}}$ for the vector obtained from $\boldsymbol{u}_r$ by replacing coordinate $d$ with $p_{r,d}$. The procedure always evaluates the FA value, regardless of the contention-resolution protocol used in the simulation, so the same notification sets are used under all contention resolution protocols.

\begin{algorithm}[H]
\caption{Deterministic rounding by conditional expectations}
\label{alg:derandomized}
\KwIn{Weights $\{w_{r,d}\}$ and acceptance probabilities $\{p_{r,d}\}$}
\KwOut{Disjoint notification sets $(S_r)_{r\in\mathcal{R}}$}

Solve \eqref{eq:LP} to obtain an optimal solution $(x^*,y^*)$\;
Set $S_r\gets\varnothing$ for every $r\in\mathcal{R}$\;
Set $u_{r,d}\gets y^*_{r,d}(1-y^*_{r,d}/3)$
for every $(r,d)$\;

\ForEach{$d\in\mathcal{D}$ in a fixed order}{
    Set $u_{r,d}\gets0$ for every $r\in\mathcal{R}$\;
    \ForEach{$r\in\mathcal{R}$}{
        Compute
        \[
        \Delta_{r,d}
        \gets
        F_r^{\text{FA}}(\boldsymbol{u}_r^{\,d\leftarrow p_{r,d}})
        -F_r^{\text{FA}}(\boldsymbol{u}_r)
        \]
    }
    Choose $r^*\in\arg\max_r\Delta_{r,d}$,
    breaking ties deterministically\;
    \If{$\Delta_{r^*,d}>0$}{
        Set $S_{r^*}\gets S_{r^*}\cup\{d\}$\;
        Set $u_{r^*,d}\gets p_{r^*,d}$\;
    }
}
\KwRet{$(S_r)_{r\in\mathcal{R}}$}\;
\end{algorithm}
\paragraph{Step 3: Trimming.}
Algorithm~\ref{alg:derandomized} does not account for the batch-size limit $U$ or the opportunity-cost threshold $\theta$, so a rounded set may contain more than $U$ drivers, or drivers whose contribution does not justify tying them up. We therefore trim each set. For each ride $r$, we order the rounded set $\mathcal{N}_r$ by decreasing $w_{r,d}$ and consider all prefixes of length at most $U$. Among the prefixes that satisfy the marginal-efficiency constraint $F_r(S)-F_r(S\setminus\{d\})\geq\theta\,p_{r,d}$ for every $d\in S$, we retain the one with the largest value of $F_r$, where $F_r$ is the valuation induced by the protocol in use (FA or BA). When $U=1$, the algorithm notifies only the highest-scoring driver in each rounded set and therefore reduces to an exclusive-dispatch rule.

\begin{remark}\label{rmk:approx}
An alternative is exhaustive trimming, which searches over all subsets of each rounded notification set rather than only its prefixes. In our simulations, this approach often produces the same solution as \textsc{Opt}. This is in line with the single-cycle experiments in \ifblind\cite{LyftI2026companion_Blind}\else\cite{LyftI2026companion_NonBlind}\fi, where the untrimmed rounded solution is already within a small fraction of the optimum on average. Prefix trimming, however, restricts the search to sets that retain the highest-scoring drivers. Although this restriction can sacrifice single-cycle scores, it produces more favorable speed--quality trade-offs in our simulations and fills out the frontier between FA and BA as shown in \Cref{fig:scatter}.
\end{remark}}




\subsection{\RNcolor{Homogeneous-Acceptance} Policies: Simulations \& Results}
\label{apx:acceptance-unaware}
\revcolor{We consider a setting in which the platform does not know (or does not use) individual drivers' acceptance probabilities. The simulator still draws each driver's response from the true probability, but the optimization model replaces the driver-specific component $\rho_d$ with a common value $\bar p = 0.4$. 

A critical observation here is the discrepancy between the population average and the effective pool average. While the arithmetic mean of the true acceptance probabilities across the entire driver population is $\approx0.55$, the effective average observed in the idle pool typically hovers around $0.4$. This gap arises from a survival bias in the system dynamics: drivers with high acceptance probabilities are matched and leave the pool rapidly, whereas drivers with low acceptance probabilities remain idle for longer durations. Consequently, ``picky'' drivers are systematically over-sampled in the available set at any given time $t$.

\begin{figure}[h]
    \centering
    \includegraphics[width=\linewidth]{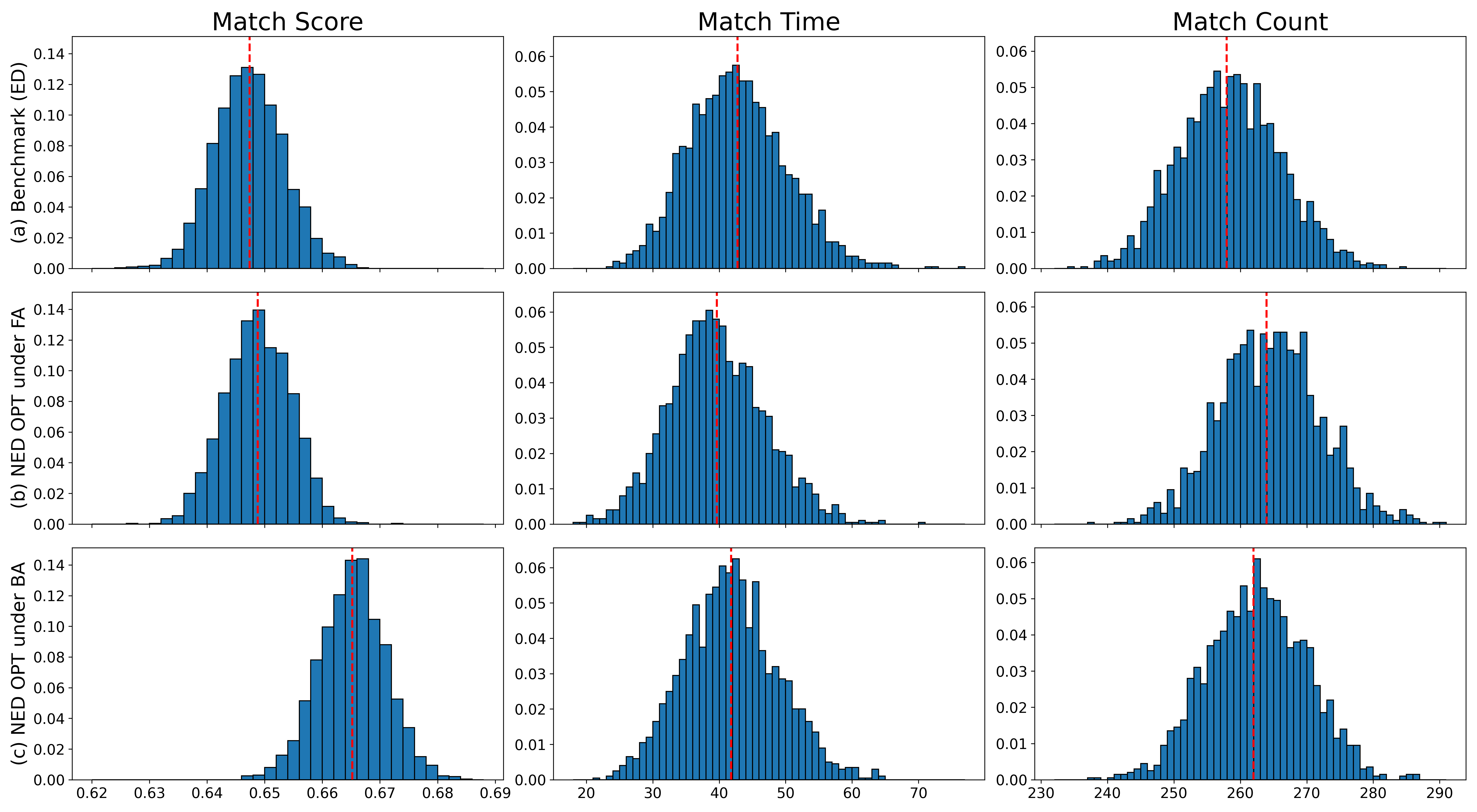}
    \caption{Exclusive dispatch that knows each driver's acceptance probability vs.\ NED OPT that
does not (static $\bar p = 0.4$), with $U=3$, $\theta=0$: (a) ED with known $p$, (b) NED OPT
under FA, (c) NED OPT under BA. Histograms, over 2000 sampled instances, of the average match
score, the average rider match time, and the number of finalized matches (fraction of
instances; dashed line = mean).}
    \label{fig:unaware}
\end{figure}

Figure~\ref{fig:unaware} compares NED OPT under this homogeneous assumption with exclusive dispatch that \emph{knows} each driver's acceptance probability. Even without driver-level information, NED outperforms the informed ED benchmark on all three metrics under both protocols: FA matches riders faster and completes more matches, and BA does so while also achieving a clearly higher match score. Moreover, losing driver-level information costs NED very little relative to the informed NED policies of \Cref{sec:simulation}. The advantage of non-exclusive notifications therefore does not rely on precise knowledge of individual acceptance behavior: notifying several drivers hedges against rejections that the platform cannot predict.}


\section{Missing Proofs \& Details of \Cref{sec:MMM}}

\revcolor{
\subsection{Sensitivity to Notification Bounds}\label{sec:MMM-sensitivity}
First, we isolate the impact of the notification upper bound $U$ on the equilibrium by holding the {opportunity-cost} threshold fixed at $\theta=0$. As illustrated in \Cref{fig:UMMM}, increasing $U$ fundamentally shifts the market's steady state. This shift is particularly pronounced when transitioning from $U=1$ to $U=2$, revealing a trade-off between match quality and market efficiency. By relaxing the notification constraint, algorithms can concurrently evaluate a broader set of options. Consequently, this significantly reduces the size of the available driver pool, leading to noticeably faster match times and higher match probabilities.

\begin{figure}[h]
    \centering
    \includegraphics[width=0.95\linewidth]{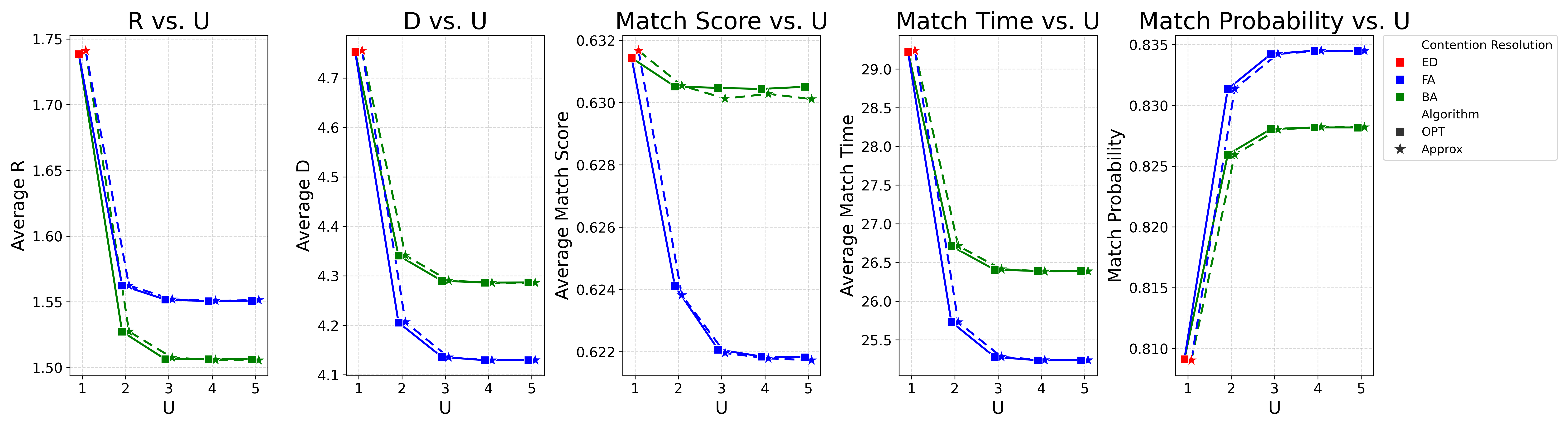}
    \caption{Sensitivity of the market equilibrium to the notification upper bound $U$ with fixed $\theta=0$.}
    \label{fig:UMMM}
\end{figure}

Next, we explore the market's sensitivity to the {opportunity-cost threshold} $\theta$, keeping the notification bound fixed at $U=3$. In \Cref{fig:ThetaMMM}, we observe that raising {$\theta$ makes the packing more conservative: a driver is notified only if their marginal contribution to the rider's welfare exceeds $\theta$ times their acceptance probability, so fewer drivers are notified per rider}. As a result, the probability of a successful match per {dispatch attempt} drops, causing unmatched riders and drivers to remain in the system longer. This accumulates into a thicker steady-state market pool but inevitably increases the average match time, highlighting the cost of waiting for higher-quality matches.

\begin{figure}[h]
    \centering
    \includegraphics[width=0.95\linewidth]{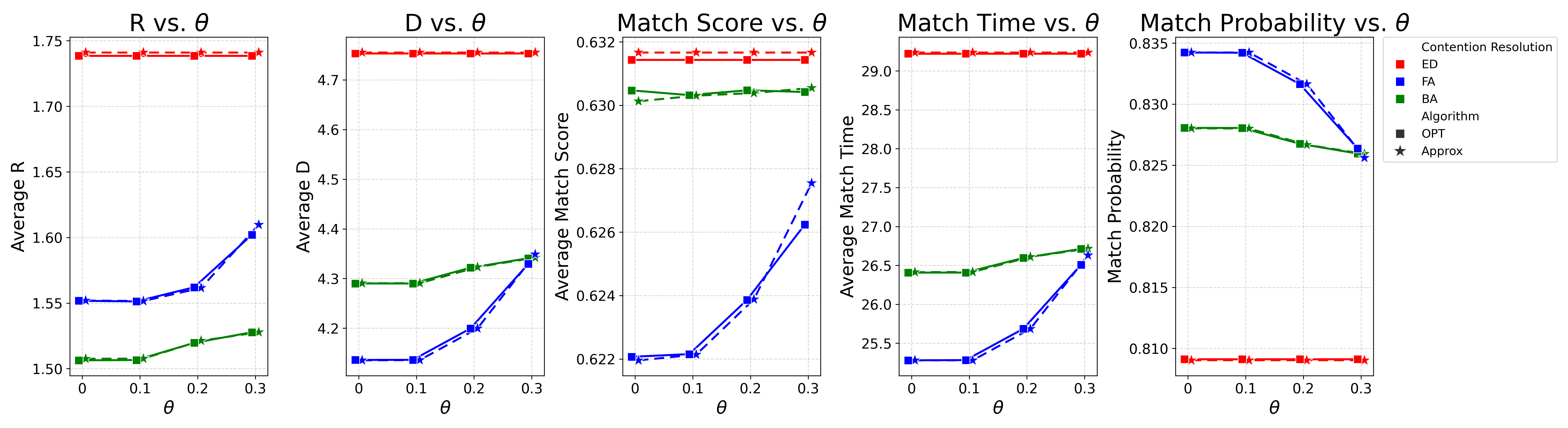}
    \caption{Sensitivity of the market equilibrium to the opportunity cost threshold  $\theta$ with fixed $U=3$.}
    \label{fig:ThetaMMM}
\end{figure}

Another point worth noting is that, in the stylized model, all drivers share the same acceptance probability. Consequently, the highest-scoring drivers are also the most valuable, and the trimmed \textsc{Approx} closely tracks \textsc{Opt}; see \Cref{rmk:approx}. This helps explain why increasing $U$ or $\theta$ changes the outcomes in the same direction under both algorithms, whereas their responses move in opposite directions in \Cref{fig:sensitivity_analysis}.

Finally, \Cref{fig:SigmaMMM} examines the impact of spatial density on the system's equilibrium. For this analysis, we fix the notification bound at $U=3$ and the {opportunity-cost threshold} at $\theta=0$, evaluating the ED and NED OPT algorithms under both FA and BA contention resolution protocols. To model varying city sizes, we sample rider and driver locations from a two-dimensional Gaussian distribution, $\mathcal{N}(0, \sigma^2)$. By modulating the standard deviation $\sigma$, we effectively expand or contract the geographic footprint of the market. The figure demonstrates how these changes in spatial dispersion dictate the final equilibrium outcomes, explicitly showing how a less dense market (a larger $\sigma$) impacts the algorithms' overall efficiency and pool sizes.

\begin{figure}[h]
    \centering
    \includegraphics[width=0.95\linewidth]{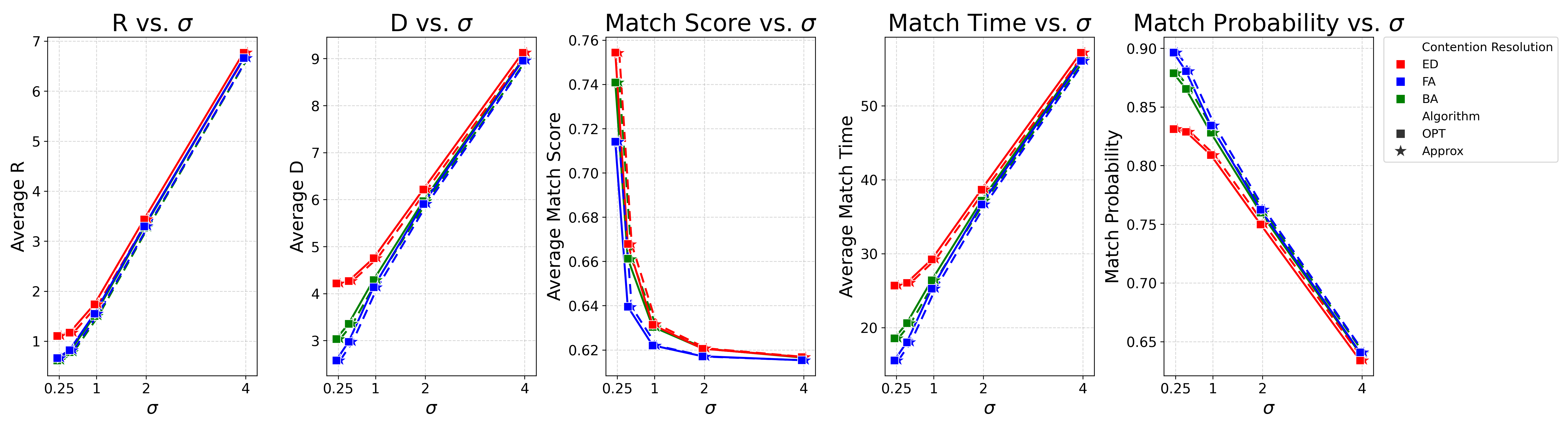}
    \caption{Sensitivity of the market equilibrium to the city density $\sigma$ with fixed $U=3$ and $\theta=0$.}
    \label{fig:SigmaMMM}
\end{figure}

As the preceding figures show, performance is substantially more sensitive to the choice of hyperparameters than to the choice between \revcolor{\textsc{Opt}} and \textsc{Approx}.
}
\subsection{Missing Proofs in \Cref{sec:MMM}}
\begin{proofof}{Proof of Proposition~\ref{prop:FA}}
By summing the first two balance equations, and adding to the third equation the first equation weighted by $\ell$, we obtain the following aggregate flow identities:
\begin{align*}
\underbrace{\riderTotalArrival}_{\textrm{Arrival}\, \textrm{Rate}} &= \underbrace{\eta\left(R_0+ \sum_{\ell=1}^U R_{\ell}\right)}_{\textrm{Churn}\, \textrm{Rate}} + \underbrace{\mu p\sum_{\ell = 1}^U \ell R_{\ell}}_{\textrm{Match}\, \textrm{Rate}}\\
\underbrace{\driverArrival}_{\textrm{Arrival}\, \textrm{Rate}} &= \underbrace{D_0\idleChurn+ \notifChurn \sum_{\ell=1}^U D_{\ell}}_{\textrm{Churn}\, \textrm{Rate}} + \underbrace{\mu p\sum_{\ell = 1}^U D_{\ell}}_{\textrm{Match}\, \textrm{Rate}}.
\end{align*}
Intuitively speaking, this line shows that each arrived driver and rider are either matched and left the platform or churned after a while. Solving for $R_{\ell}$ we get:
\begin{align*}
    R_{\ell} = \frac{R_0}{\ell\left(\mu(1-p)+\notifChurn\right)}\left[\sum_{i=\ell}^{U}\left(q_{i}\prod_{j=\ell}^i \frac{j \left(\mu(1-p)+\notifChurn\right)}{\rideChurn + j (\mu+\notifChurn)}\right)\right]
\end{align*}
Note that the above term has a Little's law interpretation: the average waiting time in state $R_{\ell}$ is $\frac{1}{\rideChurn + \ell (\mu+\notifChurn)}$, and the riders who started from $i$ drivers reach this state with rate:
\begin{align*}
    R_0 q_{i} \prod_{j=\ell+1}^i \frac{j \left(\mu(1-p)+\notifChurn\right)}{\rideChurn + j (\mu+\notifChurn)}.
\end{align*}
Therefore, we can go further and solve for $R_0$:
\begin{align*}
    R_0 = \frac{\riderTotalArrival + R_1\left(\mu(1-p)+\notifChurn\right)}{\rideChurn + \sum_{i=1}^U q_i} &= \frac{\riderTotalArrival +
    R_0\left(\sum_{i=1}^{U}q_{i} \prod_{j=1}^i \frac{j \left(\mu(1-p)+\notifChurn\right)}{\rideChurn + j (\mu+\notifChurn)}\right)}{\rideChurn + \sum_{i=1}^U q_i} \\
    &= \frac{\lambda}{\rideChurn + \sum_{i=1}^U q_i\left(1-\prod_{j=1}^i \frac{j \left(\mu(1-p)+\notifChurn\right)}{\rideChurn + j (\mu+\notifChurn)}\right)}
\end{align*}
Finally we have
\begin{align*}
    \frac{\driverTotalArrival- D_0\idleChurn}{\mu p+ \notifChurn} &= \sum_{\ell = 1}^U \ell R_{\ell}\\
    &= \frac{R_0}{\mu(1-p)+\notifChurn}\sum_{\ell = 1}^U \left( \sum_{i=\ell}^{U}q_{i} \prod_{j=\ell}^i \frac{j \left(\mu(1-p)+\notifChurn\right)}{\rideChurn + j (\mu+\notifChurn)}\right) \\
    &= \frac{\frac{\riderTotalArrival}{\mu(1-p)+\notifChurn}\sum_{\ell = 1}^U \left( \sum_{i=\ell}^{U}q_{i} \prod_{j=\ell}^i \frac{j \left(\mu(1-p)+\notifChurn\right)}{\rideChurn + j (\mu+\notifChurn)}\right)}{\rideChurn + \sum_{i=1}^U q_i\left(1-\prod_{j=1}^i \frac{j \left(\mu(1-p)+\notifChurn\right)}{\rideChurn + j (\mu+\notifChurn)}\right)}.
\end{align*}
\end{proofof}
\begin{proofof}{Proof of Proposition~\ref{prop:BA}}

By simple algebra (similar to proof of \Cref{prop:FA}) we can get aggregate equations:
\begin{align*}
\underbrace{\riderTotalArrival}_{\textrm{Arrival}\, \textrm{Rate}} &= \underbrace{\eta\left(R_0+ \sum_{\ell=1}^U R_{\ell}+\sum_{\ell=2}^U A_{\ell}\right)}_{\textrm{Churn}\,\textrm{Rate}} + \underbrace{\mu p\left(\sum_{\ell = 1}^U R_{\ell}+\sum_{\ell = 2}^U A_{\ell}\right)+\left(\mu(1-p)+\notifChurn \right) A_2}_{\textrm{Match}\, \textrm{Rate}}\\
    \underbrace{\driverTotalArrival    }_{\textrm{Arrival}\,\textrm{Rate}} &= \underbrace{D\idleChurn + \notifChurn\sum_{\ell = 1}^U D_{\ell}}_{\textrm{Churn}\,\textrm{Rate}} + \underbrace{\tilde{A}_2(\mu(1-p)+\notifChurn)+ \mu p D_1}_{\textrm{Match}\, \textrm{Rate}}.
\end{align*}
Intuitively speaking, the above two lines shows that each driver or rider is either matched or left the platform because of being idle for too long. By solving for $R_{\ell}$ we get the same formulas as FA:
\begin{align*}
    R_0 &= \frac{\lambda}{\rideChurn + \sum_{i=1}^U q_i\left(1-\prod_{j=1}^i \frac{j \left(\mu(1-p)+\notifChurn\right)}{\rideChurn + j (\mu+\notifChurn)}\right)}\\
    R_{\ell} &= \frac{R_0}{\ell\left(\mu(1-p)+\notifChurn\right)}\sum_{i=\ell}^{U}\left( q_{i} \prod_{j=\ell}^i \frac{j \left(\mu(1-p)+\notifChurn\right)}{\rideChurn + j (\mu+\notifChurn)}\right)= \frac{\frac{\lambda}{\ell\left(\mu(1-p)+\notifChurn\right)} \sum_{i=\ell}^{U}\left(q_{i} \prod_{j=\ell}^i \frac{j \left(\mu(1-p)+\notifChurn\right)}{\rideChurn + j (\mu+\notifChurn)}\right)}{\rideChurn + \sum_{i=1}^U q_i\left(1-\prod_{j=1}^i \frac{j \left(\mu(1-p)+\notifChurn\right)}{\rideChurn + j (\mu+\notifChurn)}\right)}
\end{align*}
Getting a closed form for $A_{\ell}$ is much harder, in order to do that we will use the recursions for $D_{\ell}$. Note that the flow constraint for $D_{\ell}$ will be:
\begin{align*}
    R_0\sum_{i=\ell}^Uq_i + \ell D_{\ell+1}(\mu (1-p) + \notifChurn) = D_{\ell}\left(\rideChurn + \ell (\notifChurn + \mu) \right)
\end{align*}
And we get:
\begin{align*}
    D_{\ell} = R_0\sum_{j=\ell}^U\left[\frac{\left(\prod_{i=\ell}^{j}\frac{i(\mu(1-p)+\notifChurn)}{\rideChurn + i(\notifChurn + \mu)}\right)\sum_{i=j}^Uq_i}{j(\mu(1-p)+\notifChurn)}\right]
\end{align*}
Using this we can calculate $A_i$:
\begin{align*}
    A_{\ell} =& \tilde{A}_{\ell} = D_{\ell-1}-D_{\ell}-R_{\ell-1}\\
    =& R_0\Biggr(\sum_{j=\ell-1}^U\left[\frac{\left(\prod_{i=\ell-1}^{j}\frac{i(\mu(1-p)+\notifChurn)}{\rideChurn + i(\notifChurn + \mu)}\right)\sum_{i=j}^Uq_i}{j(\mu(1-p)+\notifChurn)}\right]- \sum_{j=\ell}^U\left[\frac{\left(\prod_{i=\ell}^{j}\frac{i(\mu(1-p)+\notifChurn)}{\rideChurn + i(\notifChurn + \mu)}\right)\sum_{i=j}^Uq_i}{j(\mu(1-p)+\notifChurn)}\right] \\
    &- \frac{\sum_{i=\ell-1}^{U}\left( q_{i} \prod_{j=\ell-1}^i \frac{j \left(\mu(1-p)+\notifChurn\right)}{\rideChurn + j (\mu+\notifChurn)}\right)}{(\ell-1)\left(\mu(1-p)+\notifChurn\right)}\Biggr)\\
    =& R_0\left(\frac{\sum_{i=\ell-1}^Uq_i\left(1-\prod_{j=\ell}^i \frac{j \left(\mu(1-p)+\notifChurn\right)}{\rideChurn + j (\mu+\notifChurn)}\right)}{\rideChurn + (\ell-1)(\notifChurn + \mu)}- \sum_{j=\ell}^U\left[\frac{\frac{\rideChurn+ (\ell-1)\mu p}{\rideChurn + (\ell-1)(\notifChurn + \mu)}\left(\prod_{i=\ell}^{j}\frac{i(\mu(1-p)+\notifChurn)}{\rideChurn + i(\notifChurn + \mu)}\right)\sum_{i=j}^Uq_i}{j(\mu(1-p)+\notifChurn)}\right]\right)
\end{align*}

\end{proofof}

\begin{proofof}{Proof of Proposition~\ref{prop:MatchTime}}
    We analyze the lifecycle of a ride request as an absorbing continuous-time Markov chain using the matrices $M$ and $Q$ defined in the proposition statement. 
    
    Recall that the diagonal elements of the transient rate matrix $M$ satisfy the conservation constraint $M_{\ell \ell} = - \sum_{j \neq \ell} M_{\ell j} - (Q_{\ell 1} + Q_{\ell 2})$. Let $N$ be the fundamental matrix, where $N_{ij}$ represents the expected time spent in transient state $j$ given a start in state $i$. Standard first-step analysis yields the relation:
    \[
        N_{ij} = -\sum_{\ell\neq i} \frac{M_{i\ell}}{M_{ii}} N_{\ell j} \quad (\text{for } i \neq j), \quad
        N_{jj} = -\frac{1}{M_{jj}} - \sum_{\ell\neq j} \frac{M_{j\ell}}{M_{jj}} N_{\ell j}.
    \]
    This implies $MN = -I$, or $N = -M^{-1}$.
    
    Next, let $P_{iu}$ denote the probability that a rider starting in state $i$ is eventually absorbed into state $u$. The balance equations yield:
    \begin{align*}
        P_{iu} = \frac{Q_{iu} + \sum_{\ell \neq i} M_{i \ell}P_{\ell u}}{-M_{ii}}.
    \end{align*}
    In matrix form, this rearranges to $-MP = Q$, or equivalently:
    \begin{align}\label{eq:SuccessProb}
        P = NQ.
    \end{align}

    Finally, let $T_{iu}$ denote the expected cumulative time until absorption into state $u$, unconditioned on the event occurring (if absorption into $u$ never occurs, the contribution is zero). Recursive analysis gives:
    \begin{align*}
        T_{iu} = \frac{P_{iu}}{-M_{ii}} + \sum_{j \neq i}\frac{M_{ij}}{-M_{ii}}T_{ju}.
    \end{align*}
    Multiplying by $-M_{ii}$ and rearranging leads to the matrix equation $MT = -P$. Substituting $P=NQ$ and $M=-N^{-1}$, we obtain:
    \begin{align*}
        T = -M^{-1}P = N(NQ) = N^2 Q.
    \end{align*}
    To find the expected match time \emph{conditioned} on a successful match, we normalize the accumulated time $T_{i2}$ by the probability of success $P_{i2}$:
    \begin{align*}
        \mathbf{E}\left[\,\text{Match Time} \;\middle|\; X_0 = i,\, \text{Matched} \,\right]
        = \frac{T_{i2}}{P_{i2}}
        = \frac{\left(N^2 Q \right)_{i 2}}{ \left(N Q \right)_{i 2}}.
    \end{align*}
\end{proofof}

\end{document}